\documentclass[usenatbib,useAMS,usedcolumn]{mn2e}
\usepackage{longtable}
\usepackage{psfig}
\usepackage{epsfig}
\usepackage{graphicx}
\usepackage{times}

\def\arcm{\hbox{$^\prime$}}

\def\eg{{\it e.g.\ }}
\def\etc{{\it etc.\ }}
\def\ie{{\it i.e.\ }}

\def\deg{\hbox{$^\circ$}}
\def\spose#1{\hbox to 0pt{#1\hss}}
\def\gtsim{$\mathrel{\spose{\lower 3pt\hbox{$\sim$}}
        \raise 2.0pt\hbox{$>$}}$\thinspace}
\def\lesssim{$\mathrel{\spose{\lower 3pt\hbox{$\sim$}}
        \raise 2.0pt\hbox{$<$}}$\thinspace}
\def\simpropto{$\mathrel{\spose{\lower 3pt\hbox{$\sim$}}
        \raise 2.0pt\hbox{$\propto$}}$\thinspace}
\newcommand{\rosat}{\emph{ROSAT}}
\newcommand{\chandra}{\emph{Chandra}}
\newcommand{\xmm}{\emph{XMM-Newton}}
\newcommand{\asca}{\emph{ASCA}}
\newcommand{\einstein}{\emph{Einstein}}
\newcommand{\arcs}{\mbox{\arcm\hskip -0.1em\arcm}}
\newcommand{\Lx}{\ensuremath{L_{\mathrm{X}}}}
\newcommand{\Tx}{\ensuremath{T_{\mathrm{X}}}}
\newcommand{\Zsol}{\ensuremath{Z_{\odot}}}

\newcommand{\Msol}{\ensuremath{M_{\odot}}}
\newcommand{\LB}{\ensuremath{L_{\mathrm{B}}}}
\newcommand{\LBsol}{\ensuremath{L_{B\odot}}}

\newcommand{\LT}{\ensuremath{\mbox{\Lx :\Tx}}}

\newcommand{\Bfit}{\ensuremath{\beta_{fit}}}

\newcommand{\NH}{\ensuremath{N_{\mathrm{H}}}}

\newcommand{\s}{\ensuremath{\mbox{~s}}}
\newcommand{\ps}{\ensuremath{\s^{-1}}}
\newcommand{\km}{\ensuremath{\mbox{~km}}}
\newcommand{\Mpc}{\ensuremath{\mbox{~Mpc}}}
\newcommand{\pMpc}{\ensuremath{\Mpc^{-1}}}
\newcommand{\kmpspMpc}{\ensuremath{\km \ps \pMpc\,}}
\newcommand{\erg}{\ensuremath{\mbox{~erg}}}
\newcommand{\ergps}{\ensuremath{\erg \ps}}

\newcommand{\sas}{\textsc{sas}}
\newcommand{\Mgxii}{\mbox{{Mg{\small XII}}}}
\newcommand{\Ovii}{\mbox{{O{\small VII}}}}
\newcommand{\Fexiixiii}{\mbox{{Fe{\small XII-XIII}}}}
\newcommand{\Ho}{\ensuremath{H_\mathrm{0}}}
\newcommand{\OM}{\ensuremath{\Omega_\mathrm{M}}}
\newcommand{\OL}{\ensuremath{\Omega_\mathrm{\Lambda}}}
\newcommand{\tcool}{\ensuremath{t_{\mathrm{cool}}}}
\newcommand{\ML}{\ensuremath{\mbox{\Msol/\LBsol}}}
\newcommand{\Rth}{\ensuremath{R_{\mathrm{200}}}}
\newcommand{\Dtf}{\ensuremath{D_{\mathrm{25}}}}
\newcommand{\Lxb}{\ensuremath{L_{\ensuremath{xb}}}}
\newcommand{\LxbLb}{\ensuremath{\mbox{\Lxb :\LB}}}
\newcommand{\hsf}{\ensuremath{h_{\mathrm{75}}}}
\newcommand{\xspec}{\textsc{xspec}}
\begin{document}

\title[
An XMM-Newton observation of the galaxy group MKW 4
]
{
An XMM-Newton observation of the galaxy group MKW 4
}
\author[
Ewan O'Sullivan et al. 
]
{
E. O'Sullivan$^1$\footnotemark, J. M. Vrtilek$^1$, A. M. Read$^{2,3}$,
L.~P. David$^1$, T.~J. Ponman$^2$\\
$^1$ Harvard Smithsonian Center for Astrophysics,
60 Garden Street, Cambridge, MA 02138, USA\\
$^2$ School of Physics and Astronomy, University of Birmingham, Edgbaston,
Birmingham, B15 2TT, UK\\
$^3$ Dept. of Physics and Astronomy, Leicester University, Leicester LE1
7RH, UK \\
\\
}

\date{Accepted 2003 ?? Received 2003 ??; in original form 2003 ??}
\pagerange{\pageref{firstpage}--\pageref{lastpage}}
\def\LaTeX{L\kern-.36em\raise.3ex\hbox{a}\kern-.15em
    T\kern-.1667em\lower.7ex\hbox{E}\kern-.125emX}

\label{firstpage}

\maketitle

\begin{abstract}
  We present an X--ray study of the galaxy group or poor cluster MKW~4.
  Working with \xmm\ data we examine the distribution and properties of the
  hot gas which makes up the group halo. The inner halo shows some signs of
  structure, with circular or elliptical $\beta$ models providing a poor
  fit to the surface brightness profile. This may be evidence of large
  scale motion in the inner halo, but we do not find evidence of sharp
  fronts or edges in the emission. The temperature of the halo declines in
  the core, with deprojected spectral fits showing a central temperature of
  $\sim$1.3 keV compared to $\sim$3 keV at 100 kpc. In the central
  $\sim$30~kpc of the group multi-temperature spectral models are required
  to fit the data, but they indicate a lack of gas at low temperatures.
  Steady state cooling flow models provide poor fits to the inner regions
  of the group and the estimated cooling time of the gas is long except
  within the central dominant galaxy, NGC 4073. Abundance profiles show a
  sharp increase in the core of the group, with mean abundance rising by a
  factor of two in the centre of NGC 4073. Fitting individual elements
  shows the same trend, with high values of Fe, Si and S in the core. We
  estimate that $\sim$50\% of the Fe in the central 40~kpc was injected by
  SNIa, in agreement with previous \asca\ studies. Using our best fitting
  surface brightness and temperature models, we calculate the mass, gas
  fraction, entropy and mass-to-light ratio of the group. At 100 kpc
  ($\sim$0.1 virial radii) the total mass and gas entropy of the system
  ($\sim$2$\times$10$^{13}$\Msol\ and $\sim$300 keV cm$^2$) are quite
  comparable to those of other systems of similar temperature, but the gas
  fraction is rather low ($\sim$1\%). We conclude that MKW 4 is a fairly
  relaxed group, which has developed a strong central temperature gradient
  but not a large--scale cooling flow.
\end{abstract}


\footnotetext{Email: ejos@head-cfa.harvard.edu}

\section{Introduction}
MKW 4 is a poor cluster or rich group first identified by
\citet{Morganetal75} as part of their sample of cD galaxies in poor
clusters. It contains $\sim$50 galaxies, and is
dominated by the cD/E galaxy NGC 4073. Kinematic studies of the galaxy
population shows the galaxies to be fairly evenly distributed about their mean
redshift of $z$=0.02, with no sign of substructure
\citep{KoranyiGeller02}. There is some sign of spectral segregation, with
absorption line galaxies more commonly found toward the centre of the
group. 

NGC 4073 has an extensive globular cluster population, and shows some signs
of having undergone an interaction of some sort in the relatively recent
past. It has a counter-rotating stellar core with a significantly lower
velocity dispersion. This suggests that the core originated in a lower mass
progenitor galaxy \citep{Fisheretal95}. The rotation velocity of the
galaxy outside the core is low. The luminosity-weighted mean spectroscopic
age of the galaxy is $\sim$7.5 Gyr \citep{Terlevichforbes00}. There is no
central radio or X--ray source associated with the galaxy, suggesting that
any AGN is either dormant or weak.

MKW 4 has been the subject of X--ray studies using \einstein, \asca\ and
\rosat, all of which have shown it to be a fairly relaxed system
\citep{DellAntonioetal95,JonesForman99}, reasonably well modeled by at most
two Beta models \citep{Helsdonponman00}. The group X--ray halo has a mean
temperature of $\sim$1.7 keV, placing it at the upper end of the group \LT\ 
relation. Radial temperature profiles show that the temperature increases
from $\sim$1.3 keV in the core, to a peak above 2.5 keV at $\sim$115 kpc
and then drops back to a fairly constant temperature of $\sim$1.5 keV at
higher radii. The \asca\ data have been used to produce radial abundance
profiles \citep{Finoguenovetal00}, which show Fe, Si and S abundances which
fall from central values of 0.4-0.7\Zsol\ to 0-0.2\Zsol\ at a radius of 500
kpc. The [Si/Fe] ratio increases with radius, suggesting that in the centre
of the group SNIa are the main source of enrichment, while at higher radii
only SNII are required. Estimates based on modeling of the \rosat\ data
give the group a virial radius of $\sim$1 Mpc (Helsdon 2002, private
communication), assuming isothermality, and \Ho=75 \kmpspMpc. From the
galaxy kinematics, the mass within this radius is estimated to be
1-1.5$\times$10$^{14}$\Msol\ \citep{KoranyiGeller02}.

\begin{table}
\begin{center}
\begin{tabular}{lc}
\hline
R.A. (J2000) & 12 03 57.7 \\
Dec. (J2000) & +01 53 18 \\
Redshift & 0.02 km~s$^{-1}$ \\
Distance (\Ho=75) & 79.945 Mpc \\
1 arcmin = & 23.255 kpc \\
NGC~4073 \Dtf\ radius & 32.790 kpc\\
\hline
\end{tabular}
\end{center}
\caption{\label{tab:props} Location and scale for MKW 4.}
\end{table}

Until recently, galaxy groups and poor clusters have received relatively
little attention. From a practical viewpoint, higher mass clusters are more
easily identified in both optical and X--ray surveys, and have relatively
high X-ray luminosities which make them more rewarding targets for X-ray
observations. However, the majority of galaxies in the Universe are found
in galaxy groups \citep{Tully87}, and it is likely that a large fraction of
the total baryonic mass of the Universe is contained in such systems
\citep{ Fukugitaetal98}. Comparisons of the X--ray properties of groups
with more massive clusters show that groups cannot be treated as scaled
versions of larger systems \citep{Helsdonponman00}, and that the change in
properties occurs at $\sim$2~keV, a typical temperature for rich groups or
poor clusters such as MKW~4. Recent studies of low mass systems with \xmm\ 
and \chandra\ have also focussed on the role of feedback and metal
enrichment in their evolution
\citep{Mushotzkyetal03,Buoteetal03a,Xuetal02}. The low velocity dispersions
of less massive systems makes galaxy mergers more likely, and the
associated bursts of star formation, galaxy winds and AGN activity are
thought to be responsible for much of the enrichment of the intergalactic
medium in groups and (through the merger of groups to form larger systems)
more massive clusters. Despite these indications of the importance of these
systems, the relatively small number of groups and poor clusters which have
been studied in detail means that many of the processes which affect them
are only now being addressed. Our intention is to present an in-depth
analysis of one such system in order to understand its current state and
the factors which have affected its development.

In this work we use \xmm\ data to examine the X-ray structure of the group
and its central galaxy. Section~2 describes the observation and data
reduction techniques used. Section~3 covers our analysis of the surface
brightness distribution of the group halo, while Sections~4 and 5 detail
our spectral analysis of the data from the EPIC and RGS instruments
respectively. In Section~6 we discuss our results and conclusions.
Throughout this work we assume \Ho=75 \kmpspMpc, \OM=0.3 and \OL=0.7.
Abundances are quoted relative to the solar ratios given in
\citet{AndersGrevesse79}. This makes comparison with previous studies
simple, but we note that more recent abundance ratios
are likely to be more accurate. In particular, the Fe/H ratio given by 
\citet{GrevesseSauval98} is somewhat lower (Fe/H=3.2$\times$10$^{-5}$
compared to  4.7$\times$10$^{-5}$) than the older value used by
\citet{AndersGrevesse79}, which indicates that our Fe abundances are
underestimated by a factor of $\sim$1.4 compared to those calculated using
the more recent ratios. It is important to note this difference when
comparing our results with those of other authors.

\section{Observation and Data Reduction}
MKW 4 was observed with \xmm\ during orbit 373 (2001 December 21) in two
exposures of $\sim$16000 and $\sim$4500 seconds. The EPIC MOS and PN
instruments were operated in full frame and extended full frame modes
respectively, with the medium filter. A detailed summary of the \xmm\ 
mission and instrumentation can be found in \citet[and references
therein]{Jansenetal01}. The raw data from the EPIC instruments for the
longer of the two exposures were processed with the publicly released
version of the \xmm\ Science Analysis System (\textsc{sas v.5.3.3}), using
the \textsc{epchain} and \textsc{emchain} tasks. After filtering for bad
pixels and columns, X--ray events corresponding to patterns 0-12 for the
two MOS cameras and patterns 0-4 for the PN camera were accepted.
Investigation of the total count rate for the field revealed a short flare
at the beginning of the observation. Times when the total count rate
deviated above the mean by more than 3$\sigma$ were therefore excluded. The
effective exposure times for the MOS and PN cameras were 14.1 and 10.5
ksec, respectively. Images and spectra were extracted from the cleaned
events lists with the \sas\ task \textsc{evselect}.

The reduction of the RGS data was performed using \textsc{rgsproc-1.3.3}
and the sub-tasks within, with the position of the centre of MKW4
implicitly defined as the target of the analysis. A significant amount of
background flaring is seen in the RGS datasets, with quite a high general
rate seen in the earlier exposures (S001/S002) and a very bright,
short-duration flare seen in the later exposures (S004/S005).  The event
sets were filtered via GTI files to remove these high-background times.
This was performed within the `filter' stage of rgsproc. The remaining good
times amounted to approximately 17.0 and 16.9\,ks for the RGS1 and RGS2
instruments respectively.

\section{EPIC Imaging Analysis}
\label{sec:image}
\begin{figure*}
\centerline{\psfig{file=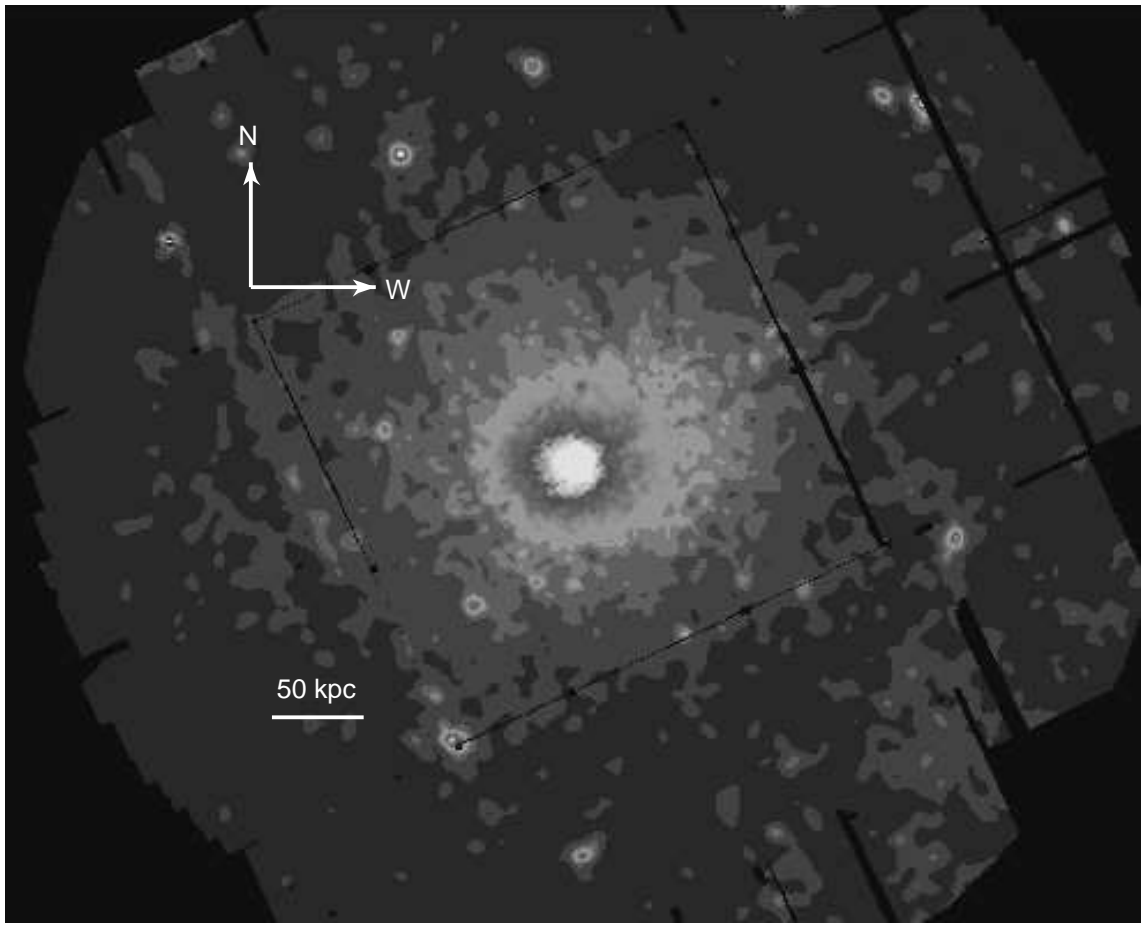,width=6in}}
\caption{\label{fig:ovly} Adaptively smoothed, mosaiced image of MKW 4,
  using data from both MOS cameras and the PN. Data from each instrument
  was corrected for vignetting and exposure, and the different effective
  areas of the instruments taken in to account during the mosaicing
  process. Smoothing was carried out using the \textsc{sas} task
  \textsc{asmooth} with a signal-to-noise ration of 10 and a maximum
  gaussian smoothing size of $\sigma$=5 pixels.}
\end{figure*}

We initially prepared an exposure and vignetting corrected mosaiced image
of the group, combining data from the PN and both MOS cameras. We
adaptively smoothed this image using the \sas\ task \textsc{asmooth}, with
a signal-to-noise ratio of 10. The resulting image, overlaid with optical
contours, can be seen in Fig~\ref{fig:ovly}.

\begin{figure}
\centerline{\psfig{file=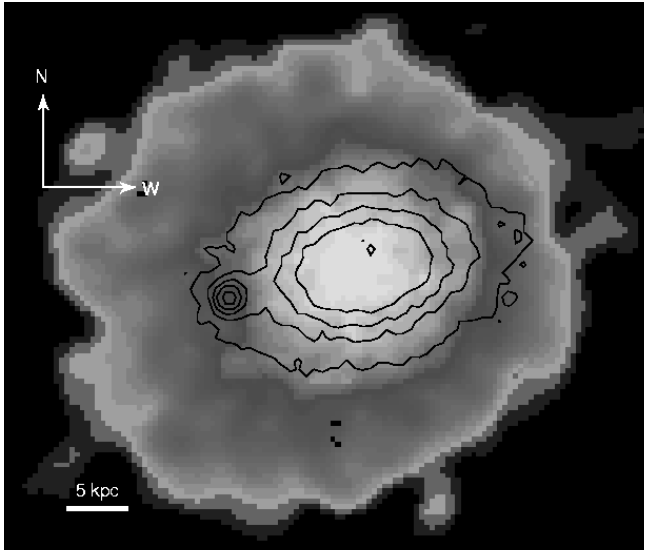,width=8.5cm,bbllx=80,bblly=210,bburx=500,bbury=570,clip=}}
\caption{\label{fig:small} MOS 1+2 image of the core of
  MKW 4 with optical contours of NGC 4073 superimposed. The data have been
  corrected and adaptively smoothed as in Fig.~\ref{fig:ovly}. The pixel size
  (1.1\arcs) and scaling have been chosen to emphasize small features in
  the X--ray emission. The point source in the optical contours at the
  eastern end of the galaxy is a foreground star.}
\end{figure}

From this image, it is clear that the group emission is extended and
relatively smooth, with no major structures evident. The dominant galaxy of
the system, NGC 4073, lies at the peak of the X--ray emission, and is
aligned roughly along the longer axis of the X--ray emission.  At small
scales, we do see some features in the X--ray emission in and around NGC
4073. Figure~\ref{fig:small} is an adaptively smoothed MOS~1+MOS~2 image of
the central part of the group, with optical contours of NGC 4073
superimposed.  A number of small linear or arcing structures can be seen,
and examination of an unsmoothed image suggests that these features are
statistically significant and likely to be real. Further investigation of
these features with \chandra\ (which observed MKW~4 with ACIS-S on
2002-11-24) might provide some clue to their nature. We do not see any
evidence of an AGN, such as a central point source, in the galaxy.

\begin{figure}
\centerline{\psfig{file=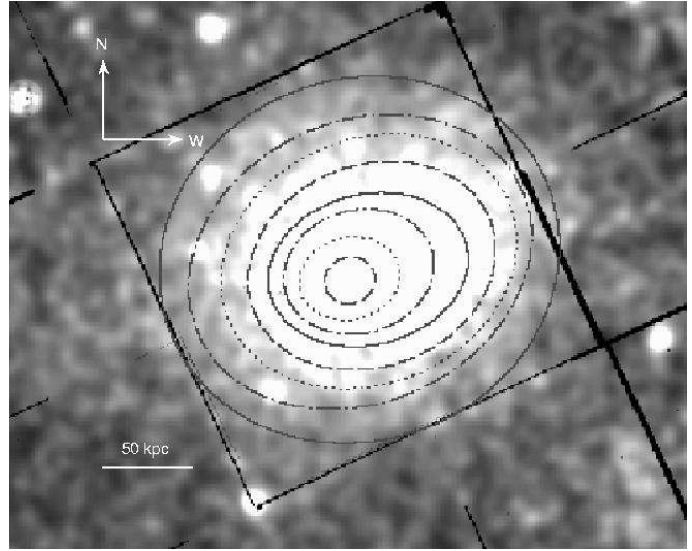, width=9cm}}
\caption{\label{fig:ellipse} Adaptively smoothed mosaic MOS image of MKW4,
  with fitted isophotal ellipses overlaid.}
\end{figure}

On larger scales, the group X--ray halo does not appear to follow a simple
elliptical form. The emission is clearly more extended to the north-west
than it is to the south-east. In order to parameterize this difference, we
decided to use the \textsc{iraf} task \textsc{ellipse} to fit isophotal
ellipses to the halo emission. To prevent the fitting being influenced by
regions of lower surface brightness caused by chip gaps on the detectors,
we used a mosaiced image containing data from the two MOS cameras only,
again adaptively smoothed to a signal-to-noise ratio of 10. The results of
this fitting can be seen in Fig~\ref{fig:ellipse}.


The ellipses fit by the task have ellipticities, position angles and
centroids which vary with radius. The group halo is clearly not well
described by any simple elliptical model. Subtraction of a model based on
the fitted ellipses left no significant residuals with the exception of
those caused by point sources in the field. 

In order to model the surface brightness distribution, we produced
images for use in the \textsc{ciao sherpa} fitting software. Data from each
camera were used to generate images in five energy bands (0.2-0.5,
0.5-2.0, 2.0-4.5, 4.5-7.5 and 7.5-12.0 keV). These were then searched for
point sources using the \sas\ sliding-cell detection task
\textsc{eboxdetect}. The source lists for each band were compared and
combined to produce a final source list for the field. Circular regions of
17\arcs\ radius were defined at each source position (except the centre of the
group halo) and excluded from all further analysis. We then prepared images
for each camera in a 0.5-3.0 keV band, chosen to maximize signal-to-noise,
binned to give 4.4\arcs\ pixels. Images were trimmed to include only a
circular region of radius 465\arcs\ centred on the peak emission. This is the
maximum radius at which we estimate that we can detect emission at more
than 3$\sigma$ significance above the background. We also used the \sas\ task
\textsc{calview} to produce on-axis PSF images for this energy band for
each camera, and binned these to the same pixel size. 

We produced background images for use with this data based on the blank-sky
datasets provided by \citet{Lumb02} and the CLOSED datasets provided by
\citet{Martyetal02}. The overall background for any EPIC observation
depends on a number of factors, including the particle flux, the position
of the region of interest on the detector and the pointing of the telescope
(owing to the possibility of soft galactic emission in the field of view).
We therefore use the blank-sky and CLOSED datasets to perform a
``double-subtraction'' of the background \citep{Arnaudetal02,Prattetal01}.
This involves identifying three main background components. The first, the
particle background, can be subtracted using the CLOSED data, scaled to match
the other datasets by comparing events recorded outside the field of view
of the cameras. Once this component is accounted for, a ``non-particle''
background can be found by subtracting the scaled particle background from
the blank-sky data and scaling the result to match the exposure of the
source observation. A final component is associated with soft galactic
emission. As the blank-sky data is unlikely to have the same contribution
from this source as the observation, comparison of source-free regions in
the two datasets should show a difference in count rates at low energies,
sometimes referred to as a ``soft excess''. A
combination of the three components produces background images for all
three cameras, binned to match the pixel scale of our source images. 

As our images of MKW~4 have low numbers of counts in many pixels, and
further binning would affect the accuracy of our determination of core
radius, we used the Cash statistic \citep{Cash79} to measure the (relative)
accuracy of our surface brightness models. This requires us to model the
background, and we use flat models, with the normalisation free to vary
independently for each camera. This will introduce some inaccuracies, as
the background is not flat. However, the variation relative to the source
is small, and using a background model allows us to effectively ignore
small variations in the background data sets.

We initially attempted to simultaneously fit these images with a variety of
model combinations (\eg single beta model, beta+beta, beta+point source,
\etc), using \textsc{sherpa}. The Cash statistic gives
no absolute measure of the goodness of the fit. However, examination of
residual images and azimuthally averaged radial profiles showed that the
fits were poor. Specifically, we were unable to fit the halo with a model
of the form used by \citet{Helsdonponman00}, an extended elliptical beta
model with an inner circular beta model representing a central cooling
region. We conclude that this difficulty is a product of the
surface-brightness distribution of the X--ray halo, and that the halo is
not well described by any of the models we used.


As the halo is not well modeled as a whole, we decided instead to model
sectors of the halo individually. We selected four sectors, covering the
major and minor axes of the halo. The SE sector was chosen to be narrower
than the other regions (50\deg\ rather than 90\deg) so as to avoid two
bright point sources which lie in the quadrant. The quadrants are defined
in Table~\ref{tab:1Dking}.

\begin{figure*}
\parbox[t]{1.0\textwidth}{\vspace{-1em}\includegraphics[width=9cm]{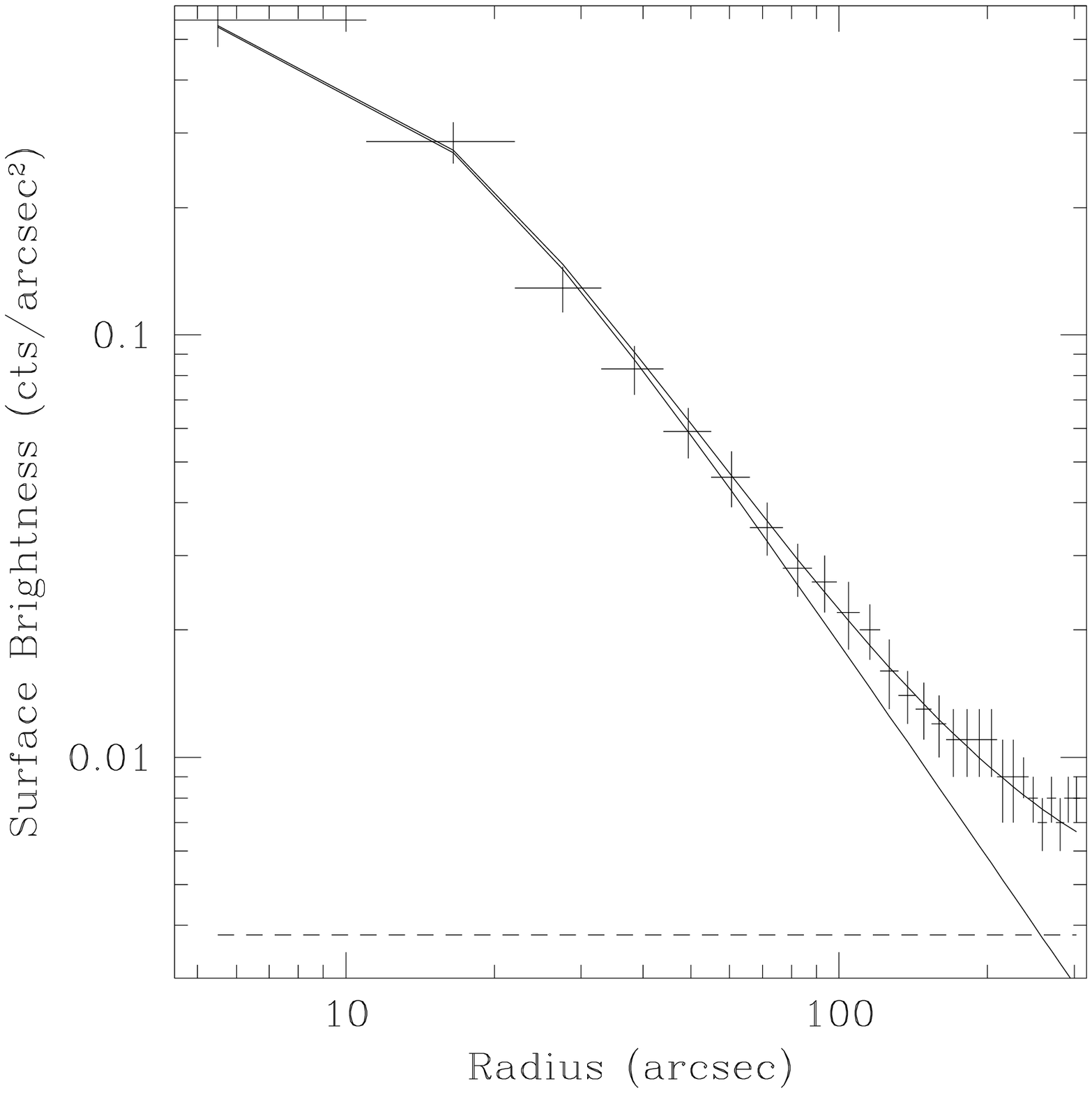}}
\parbox[t]{0.0\textwidth}{\vspace{-9.05cm}\includegraphics[width=9cm]{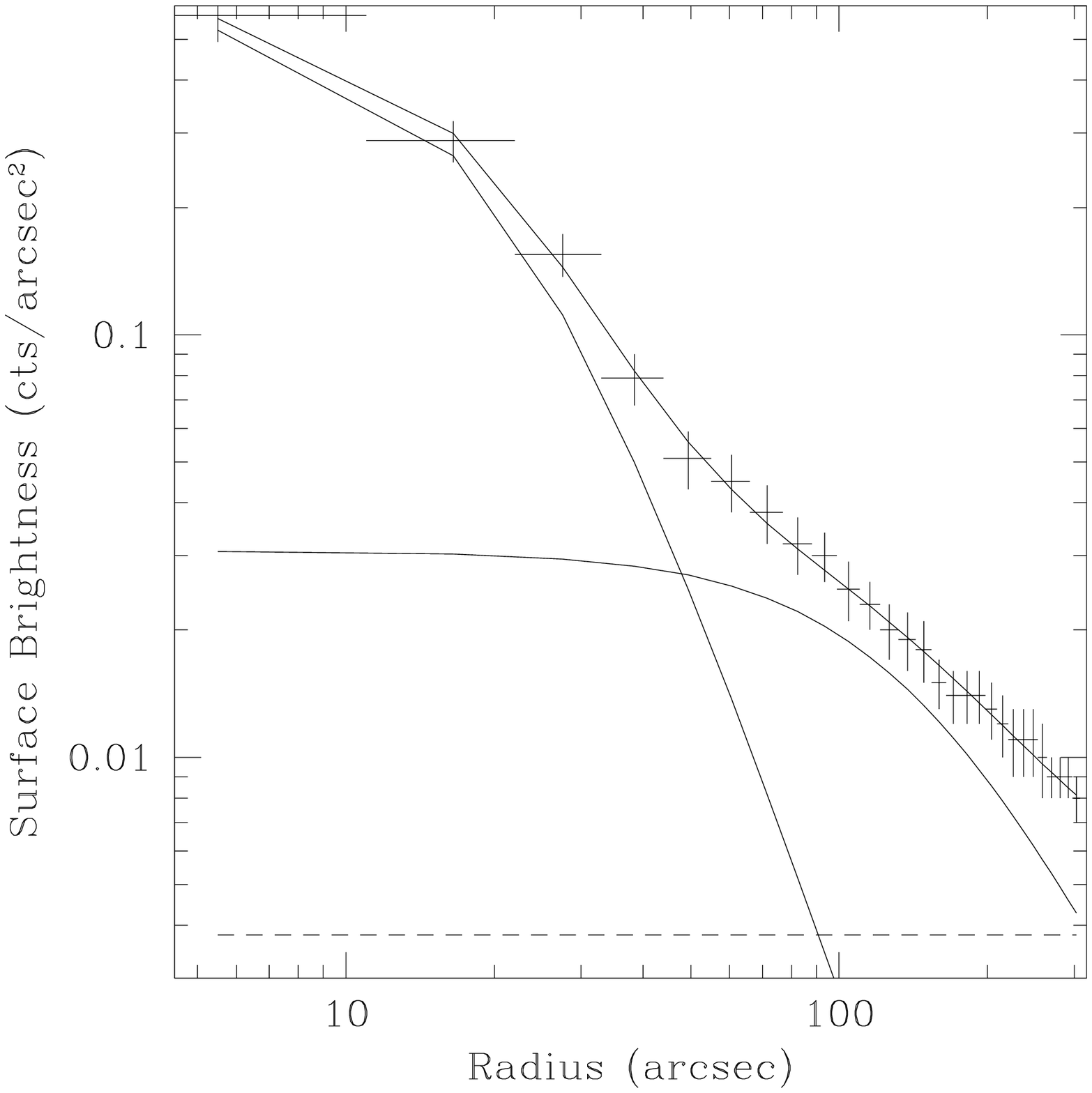}}
\parbox[t]{1.0\textwidth}{\vspace{-1cm}\includegraphics[width=9cm]{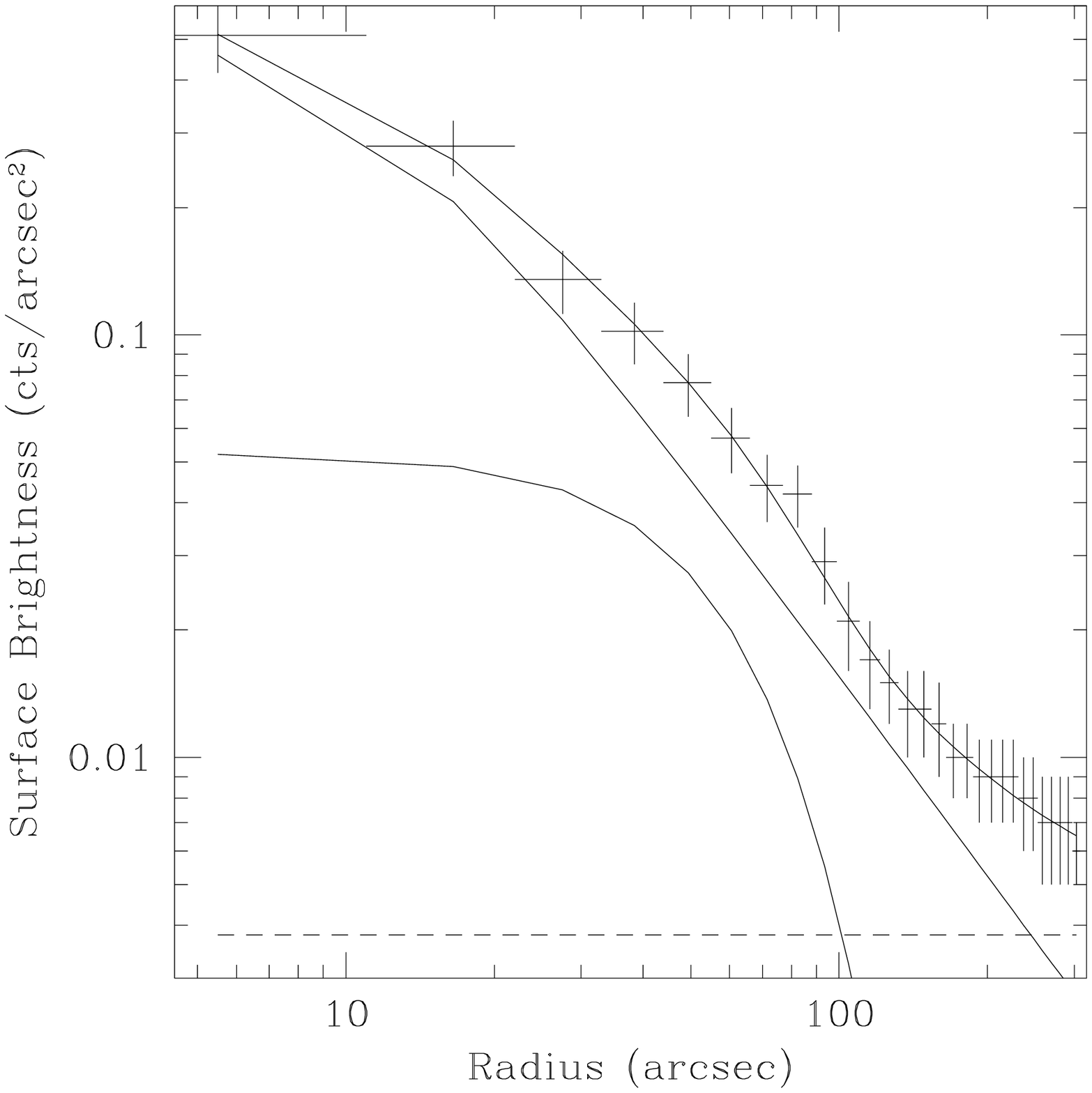}}
\parbox[t]{0.0\textwidth}{\vspace{-9.05cm}\includegraphics[width=9cm]{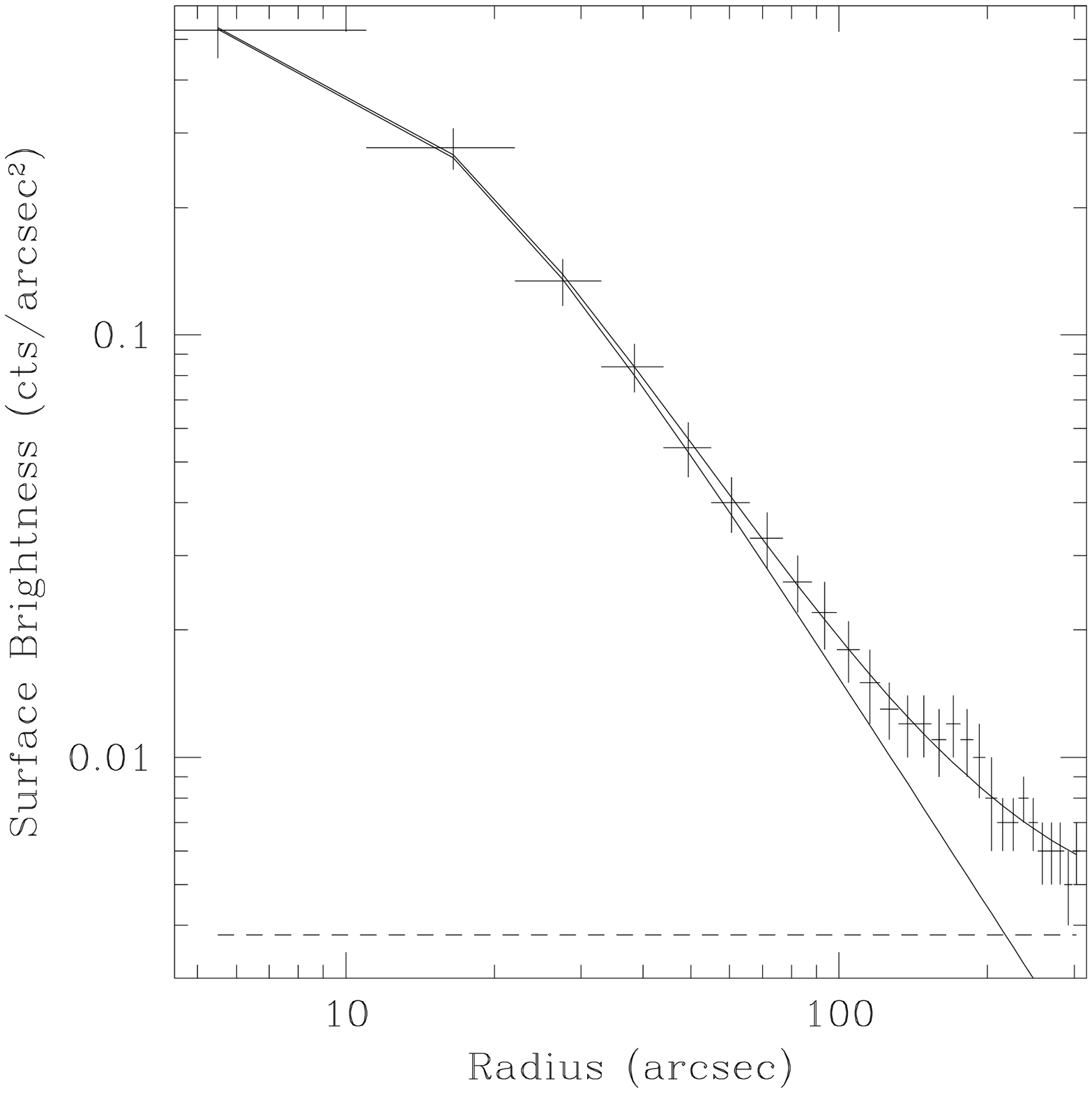}}
\caption{\label{fig:4profs}Radial surface brightness profiles of the halo
  in the four sectors described in the text. The position of each plot
  matches that of the sector, \ie they correspond to (clockwise from the
  upper left) the NE, NW, SW and SE sectors. The SE and NW sectors are not
  well modeled by a single Beta profile, while the SW and NE sectors are
  very similar. The horizontal dashed lines mark the approximate level of
  the background, while solid lines mark the models and combined
  models+background. Note that the models used for the NW and SE quadrants
  are not intended to be physically meaningful, but are the best fit to the
  data. 300\arcs\ corresponds to $\sim$116.3 kpc or $\sim$12\% of the
  virial radius.}
\end{figure*}

The NW and SE sectors are those in which we expected to find strong
deviations from a simple Beta model fit, as these lie along the axis on
which the group is most elongated. We used a mosaiced image constructed
from 0.5-3.0 keV images from the two MOS cameras to produce 1-dimensional
surface brightness profiles for each sector. Each quadrant was split into
28 annular bins of width 11\arcs, and the resulting profiles fitted with
beta models. Our intention at this stage was to test whether each quadrant
could be fitted by models with reasonable parameters, not to produce
accurate models of the halo. The fits do not take account of the effects of
the PSF, and a constant background level of 0.038 counts/arcsecond$^2$
(0.5-3.0 keV) was used in each case. Single component fits to the SE and NW
sectors produced poor results, and we added a second Beta model in each
case. However, the SW and NE sectors are well fit by a single Beta model,
and have similar values of core radius and \Bfit. Figure~\ref{fig:4profs}
shows the fits to the four quadrants, and Table~\ref{tab:1Dking}
lists the best fit parameters.

\begin{table}
\begin{center}
\begin{tabular}{l|c|cccc}
Quadrant & Angle & R$_{c,1}$ & $\beta_1$ & R$_{c,2}$ & $\beta_2$ \\
 & ($\deg$) & (kpc) & & (kpc) & \\
\hline\\[-3mm]
NE & 340-70 & 4.42$^{+1.41}_{-1.84}$ & 0.45$^{+0.01}_{-0.03}$ & - & - \\
SE & 90-140 & 3.76 & 0.43 & 127.86 & 8.43 \\
SW & 160-250 & 5.17$^{+0.84}_{-0.72}$ & 0.47$^{+0.01}_{-0.02}$ & - & - \\
NW & 250-340 & 6.77 & 0.56 & 102.31 & 0.74 \\
\end{tabular}
\end{center}
\caption{
  \label{tab:1Dking} Core radii and $\beta$-parameter
  of the 1-dimensional beta models fitted in each quadrant. Quadrants are
  defined by an angular range, where all angles are measured from north to
  east. Note that we only calculate (1$\sigma$) errors for the NE and SW
  quadrants, where the fits are statistically meaningful. 
}
\end{table}

In order to model the emission accurately in the NE \& SW quadrants, taking
into account the effect of the PSF, we again performed a 2-dimensional fit,
as described above, this time excluding all data outside the NE and SW
sectors. A single Beta model produced an adequate fit to the data, with a
core radius of 11.49$^{+0.38}_{-0.41}$\arcs\ (equivalent to
4.44$^{+0.15}_{-0.16}$kpc) and \Bfit=0.447$\pm$0.001 (1 $\sigma$ errors).
Note that the very small errors on these parameters are purely formal, and
do not take of account of systematic errors arising from (for example) our
use of a flat model to describe the background. However, the 2-dimensional
fit parameters are very similar to those found in the separate
1-dimensional fits to the NE and SW quadrants, leading us to believe that
these fits are stable and accurate.

\section{Spectroscopy with EPIC}

\begin{figure}
\centerline{\psfig{file=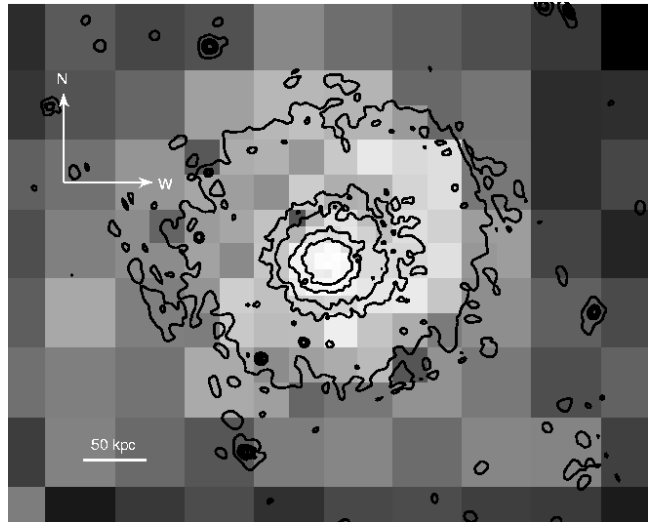,width=8.5cm}}
\caption{\label{fig:hard}Adaptively binned hardness map of MKW~4. Counts
  are binned so as to have a maximum fractional error on each bin of less
  than 15\%. Paler bins represent softer emission, darker bins are harder.
  Smoothed X-ray contours are overlaid.}
\end{figure}

In order to examine the large scale structure of the group, we produced
combined MOS and PN images in the 0.2-1.7 keV and 2.0-7.0 keV bands. We
then used these images to produce an adaptively binned hardness image
(Figure~\ref{fig:hard}), using the publicly released software of
\citet{SandersFabian01}. As expected, this showed an increase in spectral
hardness with radius, with a distribution similar to the group surface
brightness. No major deviations from this distribution were apparent,
except in areas likely to be affected by the PN chip gaps. We conclude from
this result that the temperature structure of the group is fairly simple,
and move on to spectral fitting.

The events lists used in creating the source and background
spectra were filtered using the \sas\ command \textsc{evselect} with the
expression `(FLAG == 0)' to remove all events potentially contaminated by
bad pixels and columns. All data within 17\arcs\ of point sources were also
removed, excluding the false source detection for the core of NGC 4073. We
allowed the use of both single and double events in the PN spectra, and
single, double, triple and quadruple events in the MOS spectra. The
background spectra were created by a method analogous to that used for the
spatial background files, described in the Section~\ref{sec:image}. The
same region used to extract the source spectrum was used to extract a
spectrum from the blank sky background data and scaled to match the
exposure time of the source spectrum. A 'soft excess' spectrum was then
produced based on a large radius source-free region of the source dataset,
and added to the background after appropriate scaling for area. Response
files for each spectrum were generated using the \textsc{arfgen} and
\textsc{rmfgen} commands in \sas. The data from each camera were binned to
a minimum of 20 counts per bin, and simultaneous fitting of all three
cameras was carried out in \xspec.

We extracted spectra from a circular region of 2\arcm\ (46.5 kpc) radius,
centred on the point of highest surface brightness. This region was chosen
to maximize the signal-to-noise ratio of the spectra, and to include the
high surface brightness core of the group. We ignored data below
0.2 keV and above 8.0 keV so as to minimise the use of data where the
calibration is uncertain. We then fit a series of models to the data,
starting with simple single temperature MEKAL
\citep{Liedahletal95,Kaastramewe93} and APEC \citep{Smithetal01} models and
moving on to multi-temperature and cooling flow models.
Table~\ref{tab:models} shows the results of the spectral fits.  Single
temperature fits did not successfully model the spectra, as the continuum
above 3 keV was significantly underestimated. Two-temperature fits do not
constrain the temperature of the hotter component well, and have poorly
constrained metallicities unless the metallicity is equal in the two
components, as is the case in the APEC+APEC model shown in the table. The
hot component is more successfully modeled using a powerlaw or high
temperature bremsstrahlung component, although these may also be poorly
constrained.  The best fit was attained using an APEC+Powerlaw model in
which O, Si, S and Fe were allowed to vary independently and all other
metals varied collectively. The MKCFLOW cooling flow model was a very poor
fit to the data, and allowing individual elemental abundances to vary
(VMCFLOW) did not lead to a satisfactory fit. The CEMEKL (and CEVMKL)
model, which models a multi-temperature gas where the variation of emission
with temperature follows a powerlaw, provided a much better ``cooling gas''
fit. We note that the APEC model gave a slightly better fit in most cases
than an equivalent MEKAL model, but that the difference in fit statistic
and parameters was generally fairly small.  Replacing the APEC model in the
fits shown by a MEKAL model generally did not affect the results
significantly. We also note that although the fits shown have hydrogen
column free to vary; holding it fixed at the galactic column
(1.89$\times$10$^{20}$ cm$^{-2}$) generally affected the results by only a
few percent.

\begin{table*}
\begin{tabular}{lcccccc}
Model & \NH\ & \multicolumn{3}{c}{Parameters} & red. $\chi^2_{\nu}$ & d.o.f \\
\hline\\[-3mm]
MEKAL & 4.04$\pm$0.34 & kT=1.73$\pm$0.03 & Z=0.76$^{+0.06}_{-0.05}$ & &
1.554 & 802 \\[1mm]
APEC & 4.03$^{+0.33}_{-0.32}$ & kT=1.78$\pm$0.03 & Z=0.81$^{+0.06}_{-0.05}$ & &
1.543 & 802 \\[1mm]
APEC+Powerlaw & 3.87$^{+0.50}_{-0.37}$ & kT=1.60$\pm$0.02 &
Z=0.68$\pm$0.06 & $\Gamma$=0.85$^{+0.24}_{-0.60}$ & 1.141 & 800
\\[1mm]
APEC+APEC & 3.21$^{+0.35}_{-0.36}$ & kT=1.58$\pm$0.02 &
Z=0.83$^{+0.07}_{-0.06}$ & kT=64 & 1.157 & 800 \\[1mm]
VAPEC+Powerlaw & 3.65$^{+0.44}_{-0.35}$ & kT=1.60$\pm$0.02 &
Z$_{avg}$=0.70$^{+0.17}_{-0.15}$ & $\Gamma$=0.52$^{+0.46}_{-0.50}$ & 1.074 & 796 \\[1mm]
 & & & Fe=0.56$\pm$0.05 & & & \\[1mm]
 & & & Si=0.78$\pm$0.12 & & & \\[1mm]
 & & & S=0.69$^{+0.13}_{-0.14}$ & & & \\[1mm]
 & & & O=0.29$\pm$0.11 & & & \\[1mm]
MKCFLOW & 2.12 & T$_{max}$=3.71 & \.{M}=12.12 & T$_{min}$=0.35 & 3.103 & 801 \\[1mm]
 & & & Z=2.76 & & & \\
VMCFLOW & 1.61 & T$_{max}$=3.72 & \.{M}=12.50 & T$_{min}$=0.45 & 2.342 & 797 \\
 & & & Z$_{avg}$=4.00 & & & \\
 & & & Fe=2.557 & & & \\
 & & & Si=4.57 & & & \\
 & & & S=3.23 & & & \\
 & & & O=1.25 & & & \\[1mm]
CEMEKL & 2.92$^{+0.34}_{-0.34}$ & T$_{max}$=2.24$\pm$0.07 & Z=1.13$^{+0.09}_{-0.08}$ & $\alpha$=4.23$^{+0.28}_{-0.25}$ & 1.300 & 801 \\[1mm]
CEVMKL & 2.29$^{+0.36}_{-0.35}$ & T$_{max}$=2.31$\pm$0.07 &
Z$_{avg}$=1.37$^{+0.51}_{-0.36}$ & $\alpha$=3.96$^{+0.27}_{-0.23}$ & 1.190 & 797 \\[1mm]
 & & & Fe=1.26$^{+0.27}_{-0.19}$ & & & \\[1mm]
 & & & Si=1.66$^{+0.43}_{-0.31}$ & & & \\[1mm]
 & & & S=1.28$^{+0.37}_{-0.28}$ & & & \\[1mm]
 & & & O=0.58$^{+0.23}_{-0.17}$ & & & \\[1mm]
\end{tabular}
\caption{\label{tab:models} Model fits to the spectrum extracted from the
  central 2\arcm\ of the group. All models included a WABS galactic absorption
  component \protect\citep{MorrisonMcCammon83}, given in the table in units
  of 10$^{20}$ cm$^{-2}$. The mean galactic value of \NH\ is 1.89$\times$10$^{20}$ cm$^{-2}$ \protect\citep{DickeyLockman90} 
  . Temperatures are in keV, metallicities are in Solar units, and
  $\Gamma$ is the photon index of any powerlaw component. Where individual
  elements are allowed to vary independently, the remaining elemental
  abundances are tied and fit as a single parameter, the value of which is
  given by Z$_{avg}$. In the CEMEKL and CEVMKL
  models, emission measure follows a power law in temperature, so that at a
  given temperature it is proportional to (T/T$_{max}$)$^\alpha$. 90\%
  errors are quoted for all parameters except those in the MKCFLOW and
  VMCFLOW models, which were too poor a fit to the data for errors to be
  calculated, and the high temperature component of the APEC+APEC model,
  where kT was essentially unconstrained.} 
\end{table*} 

We also carried out similar fits to a smaller region of radius 75\arcs,
again centred on the peak in surface brightness. These fits produced
similar results to those for the large region, with single-temperature
models providing poor fits and two- or multi-temperature models
favoured. We again found that the best fit was produced by an APEC+Powerlaw
model, with somewhat higher abundance and lower temperature, as
expected. Cooling flow models were poorer fits in this region, and the
difference in fit quality between the CEVMKL model and the best fitting
model was more marked. This suggests that even in the central regions, a
cooling flow model may not be the best description of the state of the gas. 

\subsection{Radial profiles}
\begin{figure*}
\parbox[t]{1.0\textwidth}{\vspace{-1em}\includegraphics[width=9cm]{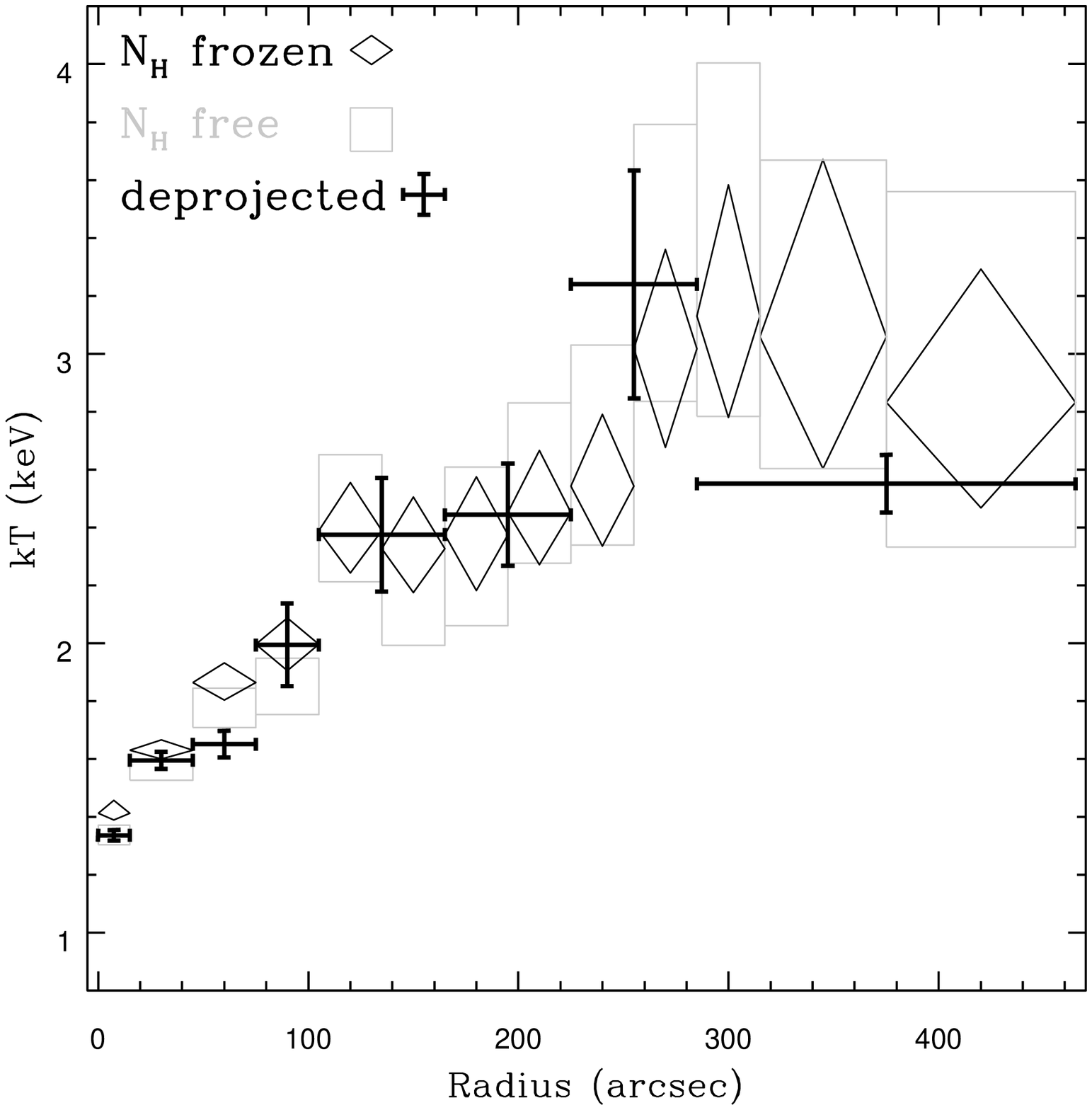}}
\parbox[t]{0.0\textwidth}{\vspace{-12.05cm}\includegraphics[width=9cm]{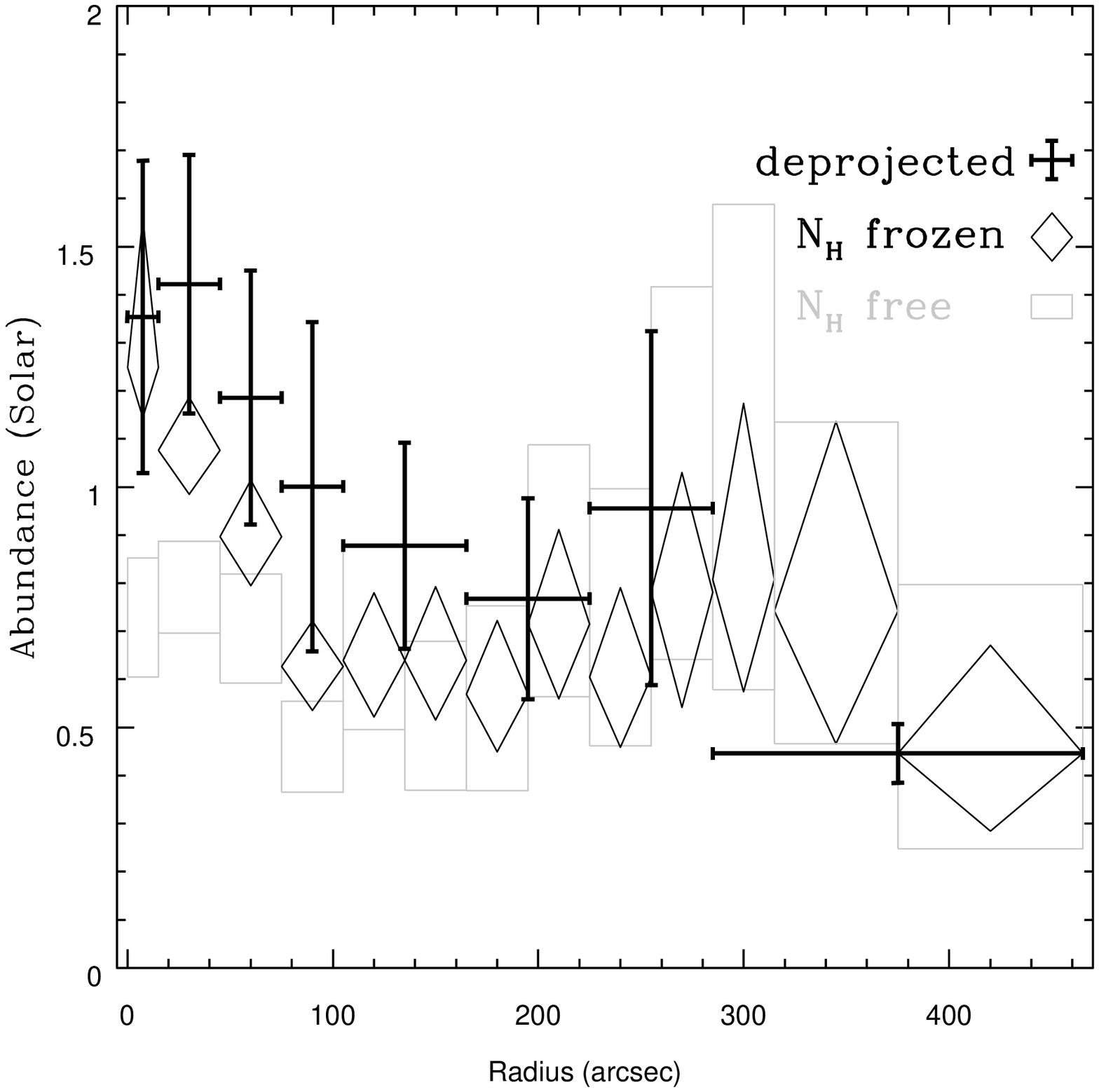}}
\parbox[t]{1.0\textwidth}{\vspace{-4cm}\includegraphics[width=9cm]{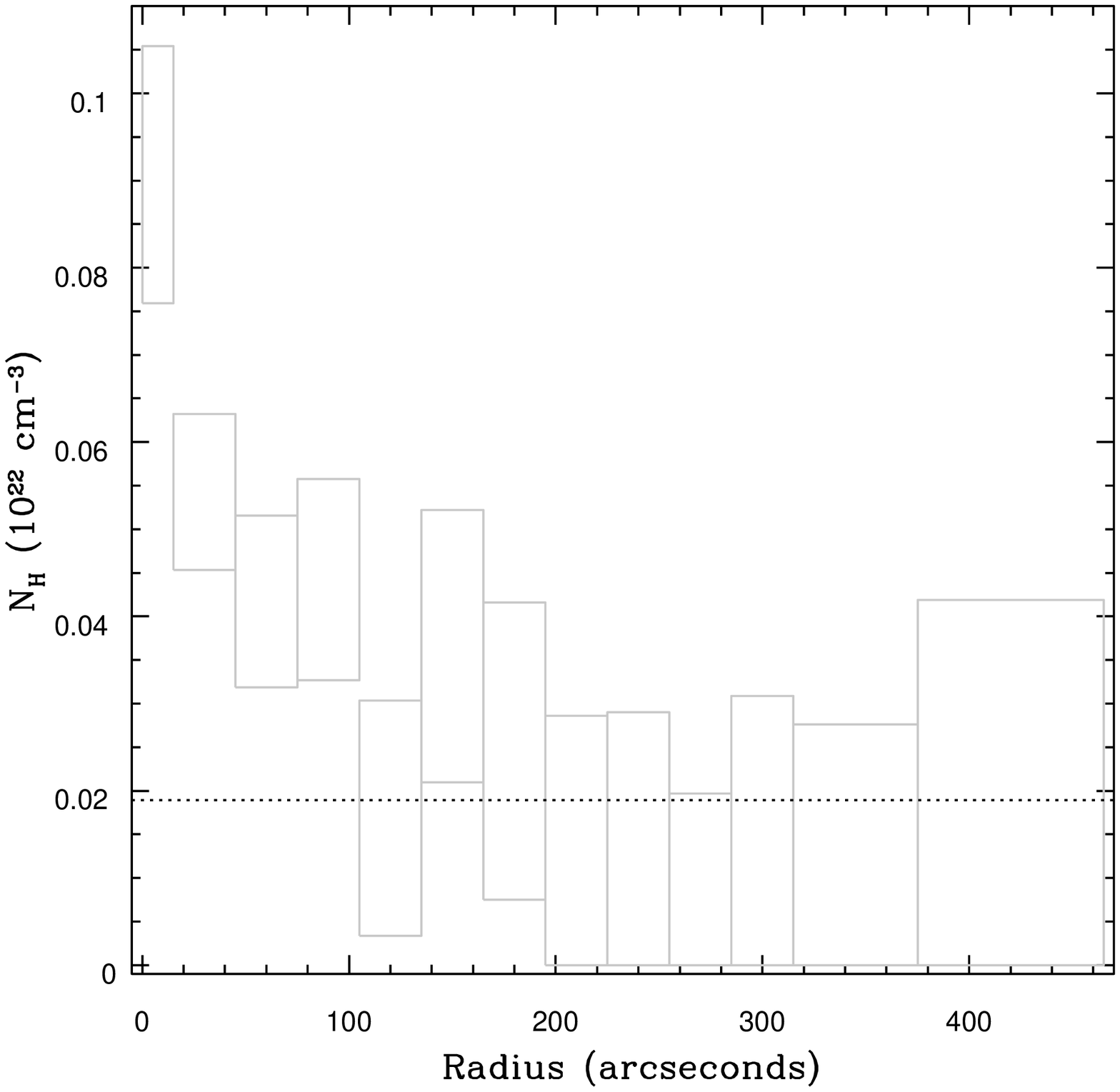}}
\parbox[t]{0.0\textwidth}{\vspace{-12.05cm}\includegraphics[width=9cm]{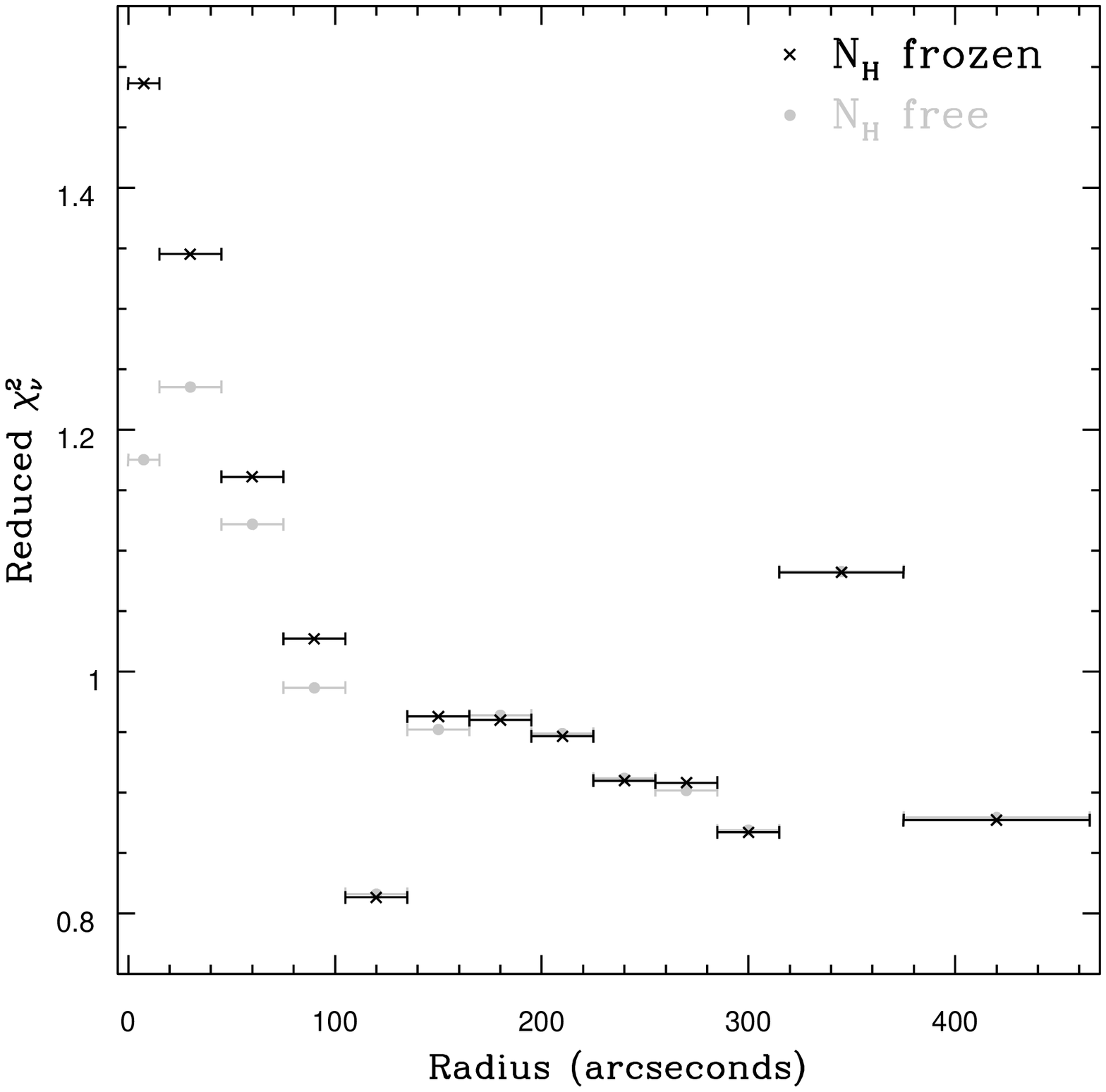}}
\vspace{-3cm}
\caption{\label{fig:Tprof} Spectral profiles for MKW~4. The upper panels
  show projected and deprojected temperature and metal abundance profiles,
  while the lower panels show the hydrogen column and fit statistic
  associated with each bin in the projected fits. In the temperature and
  abundance plots, diamonds show 90\% limits on projected fits in each
  annulus with \NH\ frozen at the galactic value. Grey rectangles show the
  limits on fits with the hydrogen column free to vary, and deprojected
  values are marked by bold crosses. The \NH\ plot has the galactic
  absorption value marked at as a dotted line, while the rectangles show
  the 90\% limits for each annulus when \NH\ is fit.  In the fit statistic
  plot, the reduced $\chi_\nu^2$ of the fit with \NH\ frozen is shown by
  crosses, and with grey circles for fits with \NH\ free. Error bars
  indicate the width of each bin.}

\end{figure*}

Figure~\ref{fig:Tprof} shows projected
temperature, metal abundance and absorption profiles for the group. Spectra
were extracted in circular annuli of varying width, and fit using an
absorbed MEKAL model in a 0.4-4.0 keV
energy band. We initially fit with hydrogen column frozen at the mean
galactic value, and then refit with the
column free to vary. Annuli were chosen to have a minimum width of 30\arcs,
so as to avoid under-sampling the PSF (at least in the core), and to have at
least 6000 counts per annulus.

The temperature profile shows a fairly smooth increase in temperature with
radius, with a possible peak at \mbox{$\sim$300\arcs}, and either a falling
or constant temperature outside this radius. This is consistent with
\rosat\ results which show the same peak, and fall to a temperature of 1-2
keV at larger radii \citep{Helsdonponman00}. The temperature profile is not
significantly affected by freeing \NH, except in the two central bins,
where kT drops slightly. Metal abundance, on the other hand, is strongly
affected by hydrogen column. If \NH\ is frozen at the galactic value, the
abundance is fairly constant (within errors) outside 75\arcs, but shows a
sharp rise to $\sim$1.25\Zsol\ in the central bins. If \NH\ is allowed to
vary, the central metal abundance is comparable to the values
measured at higher radii, but the central absorption column rises to
$\sim$4.5 times the galactic value. Although the calibration of the EPIC
instruments is imperfect at low energies, and could in principle affect the
fitted value of \NH, we note that our lower cutoff energy of 0.4 keV should
allow us to avoid major calibration uncertainties. We therefore assume that
the high values of \NH\ do not arise from poor calibration.

The quality of the fit is acceptable in all but four of the bins. In Bin 12
the poor fit seems to arise from a poorer than expected signal-to-noise
ratio for the PN spectrum. The reason for this is unclear, as there are
$\sim$4400 source counts in the PN spectrum (compared to $\sim$4100 for the
combined MOS1+2 spectra), and inspection of region from which the spectrum
is extracted suggests that we have thoroughly removed all point sources.
Fitting data from the two MOS cameras without the PN results in fits which
agree reasonably well with those for all three cameras, but with somewhat
larger errors (T=3.06$^{+0.76}_{-0.56}$, Z=0.57$^{+0.45}_{-0.28}$). The
reduced $\chi^2$ for the fit is 1.016, with 98 degrees of freedom.  The
central three bins are also rather poorly modeled by the single-temperature
model, though there is a significant improvement when hydrogen column is
allowed to vary. The poor fits in these bins seem to be an indication of
the complex nature of the emission. To test the accuracy of the single
temperature fits we tried fitting more complex models in each of the three
bins. In each case we tried MKCFLOW, CEVMKL, MEKAL+Powerlaw and MEKAL+MEKAL
fits, in the last case tying the metal abundances of the two components
together and holding the higher temperature at 2.3~keV. This temperature
was chosen to match that in the bins between 100-200\arcs. We used the
F-statistic to test for significant improvements over the single
temperature model. In all cases the CEVMKL model was a better fit that the
single temperature model, while the MKCFLOW was always a poor fit. The
powerlaw slope of the emission measure used in the CEVMKL model was very
steep in all three bins ($\alpha$ in the range 4.5-6.0, with 90\% upper
limits as high as 9.4), indicating the lack of gas at very cool
temperatures. Fits in bins 1 and 3 were also improved by using a 2-T MEKAL
model, while the MEKAL+Powerlaw was poorly constrained in all cases,
providing no superior fits. The low temperature components of the 2-T
models had temperatures of 1.17$^{+0.05}_{-0.06}$ keV,
1.43$^{+0.12}_{-0.09}$ keV and 1.22$^{+0.25}_{-0.14}$ keV for bins 1, 2 and
3 respectively (90\% errors). These temperatures again argue for a lack of
gas at low temperatures. We also note that the abundances derived, while
more poorly constrained, agree well with those found for the single
temperature models.

\begin{figure*}
\vspace{-1cm}
\centerline{\epsfig{file=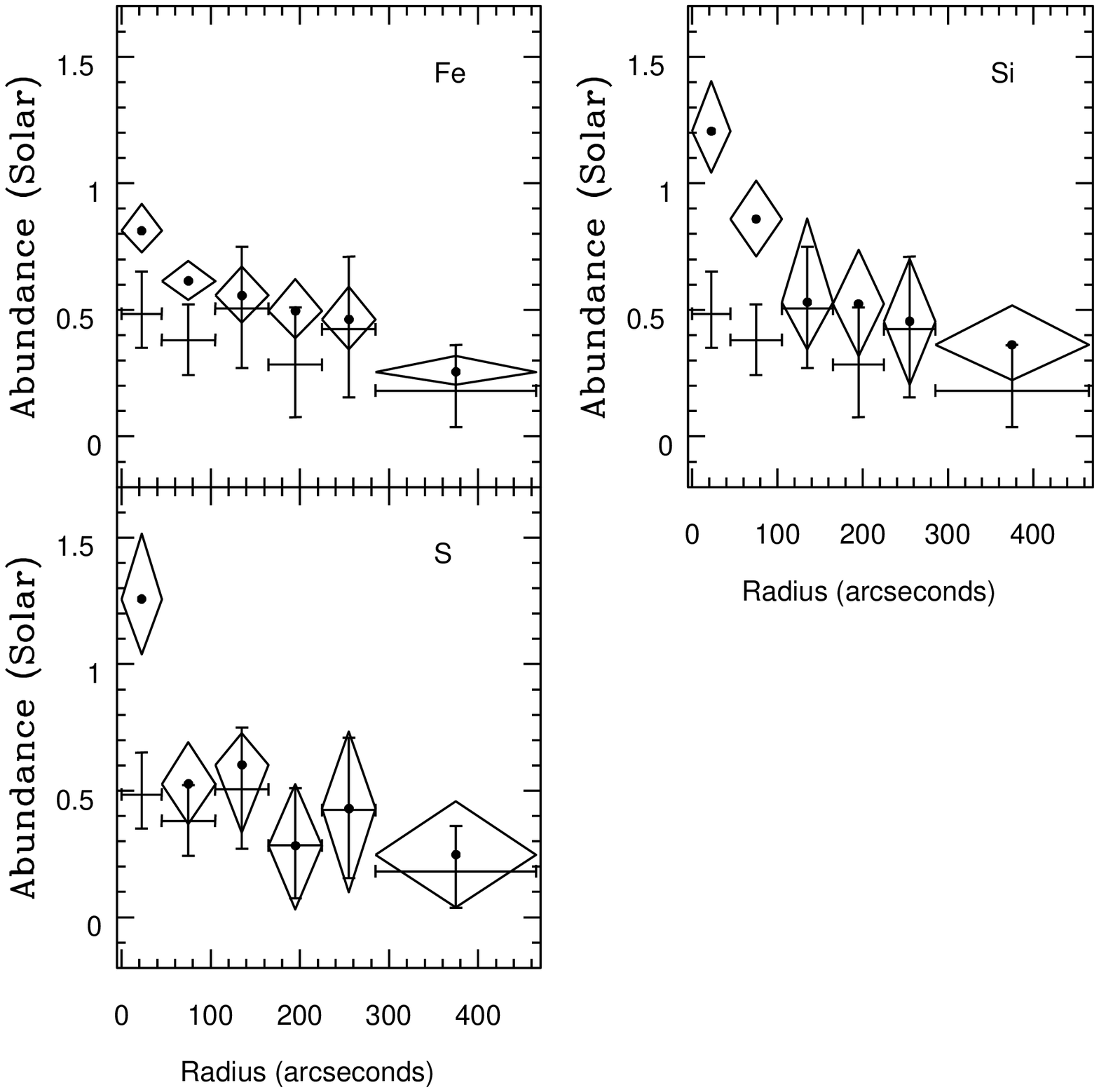, width=7.5in, height=9in}}
\vspace{-6cm}
\caption{\label{fig:Metals} Radial abundance profiles for Fe, Si and S. In
  each plot, diamonds show the 90\% error on the metal in question, as
  labeled on the plot. The crossed error bars show the fit to all the
  remaining metals, again with 90\% errors. \NH\ was fixed at the galactic
  value in all fits.}
\end{figure*}

Previous work using \asca\ has shown that the relative abundances of Fe and
Si change with radius in MKW4 \citep{Finoguenovetal00}. Although our data
do not extend to as large a radius as the \asca\ study considers, it is
ideally suited to examining the variation of individual metals near the core
of the group, and within NGC 4073. Using the VMEKAL model in \xspec\
we are able to fit Fe, Si and S individually, as well as the abundance of
remaining metals and the temperature in each bin. We use larger annular
bins to increase the signal-to-noise ratio. Figure~\ref{fig:Metals} shows
abundance profiles for Fe, Si and S, shown with reference to the profile
for all other metals combined. The hydrogen column was held at the galactic
value for these fits. It is clear that the individual metals are consistent
with the general fit at large radii, but that Fe and Si have higher
abundances in the central two bins. The S abundance is higher than the norm in
the innermost bin. It should be noted that the two inner bins cover the
region in which NGC 4073 lies, and are therefore likely to contain the most
complex emission. We are modeling this with only a single temperature
plasma, and fitted parameters are likely to be affected by this
over-simplification.

As the temperature profile is fairly smooth, with no major discontinuities,
we have also attempted to model the deprojected temperature profile of the
group. We again model the emission (in \xspec) as an absorbed MEKAL
plasma, but use the \textsc{projct} mixing model to take into account the
superposition of multi-temperature gas along the line of sight. We assume
the halo is spherical. The process of deprojection effectively lowers the
numbers of counts in the central bins, so we extract our spectra from
slightly wider annuli to increase the signal-to-noise ratio. We again fit
each spectrum in the 0.4-4 keV band, and hold the absorption column at the
galactic value. We fit each annulus separately, starting with the
outermost. Once the temperature and abundance of this annulus are fit, we
freeze those parameters and move in to the next annulus. In this manner we
fit all eight annuli, with results shown in the upper panels of Figure~\ref{fig:Tprof}. 

The deprojected abundance profile is consistent with the projected profile
in all bins except the outermost. This bin is very wide, and while it is
consistent with the projected abundance value from 375\arcs\ outward, it is
significantly lower than the projected abundance from 285\arcs\ to
375\arcs. However, the trend for abundance to decrease with radius is clear
in both projected and deprojected data. The deprojected temperature profile
is also generally consistent with the projected profile, but shows
significant differences in two of its eight bins. Although the difference
in the 8$^{th}$ (outermost) bin is likely to be caused by the lack of
counts available at higher radii and the width of the bin, the difference
in the 3$^{rd}$ bin, though small, is likely real. This may be a product of
the structure in the halo and our use of circular annuli, as this bin covers
the radii at which the surface brightness profiles of the northeast and
southwest sectors deviate from a standard beta model. However, in general
the consistency between projected and deprojected profiles suggests that
the group has a strong peak in surface brightness towards the core, so that
emission at any 2-dimensional radius is dominated by the gas within that
3-dimensional radius, not by the outer parts of the halo along the line of
sight.

\section{Spectroscopy with RGS}

The nuclear region of MKW4 is quite bright and, though quite extended, is
not too extended to allow a study of the emission with the high spectral
resolution RGS instruments \citep{DenHerderetal01} on board XMM-Newton. The
spectral range of the RGS instruments is in fact well suited to the study
of the thermodynamic properties of the hot gas within cool clusters and
groups.  The default RGS pipeline assumes that the target source is
point-like. In our analysis, as MKW4 is extended, we assumed a value of
96\% for the fraction of the cross-dispersion PSF sampled in the spectral
extraction (as opposed to the default value of 90\%). Indeed the low
surface brightness emission from MKW4 covers the RGS detector to such an
extent that the extraction of any reasonably representative non-cluster
background spectra was severely hampered. The background spectra were
extracted from areas beyond 98\% of the cross-dispersion PSF (any larger
resulted in background spectra with unreasonably low signal-to-noise
statistics).

\begin{figure*}
\centerline{\psfig{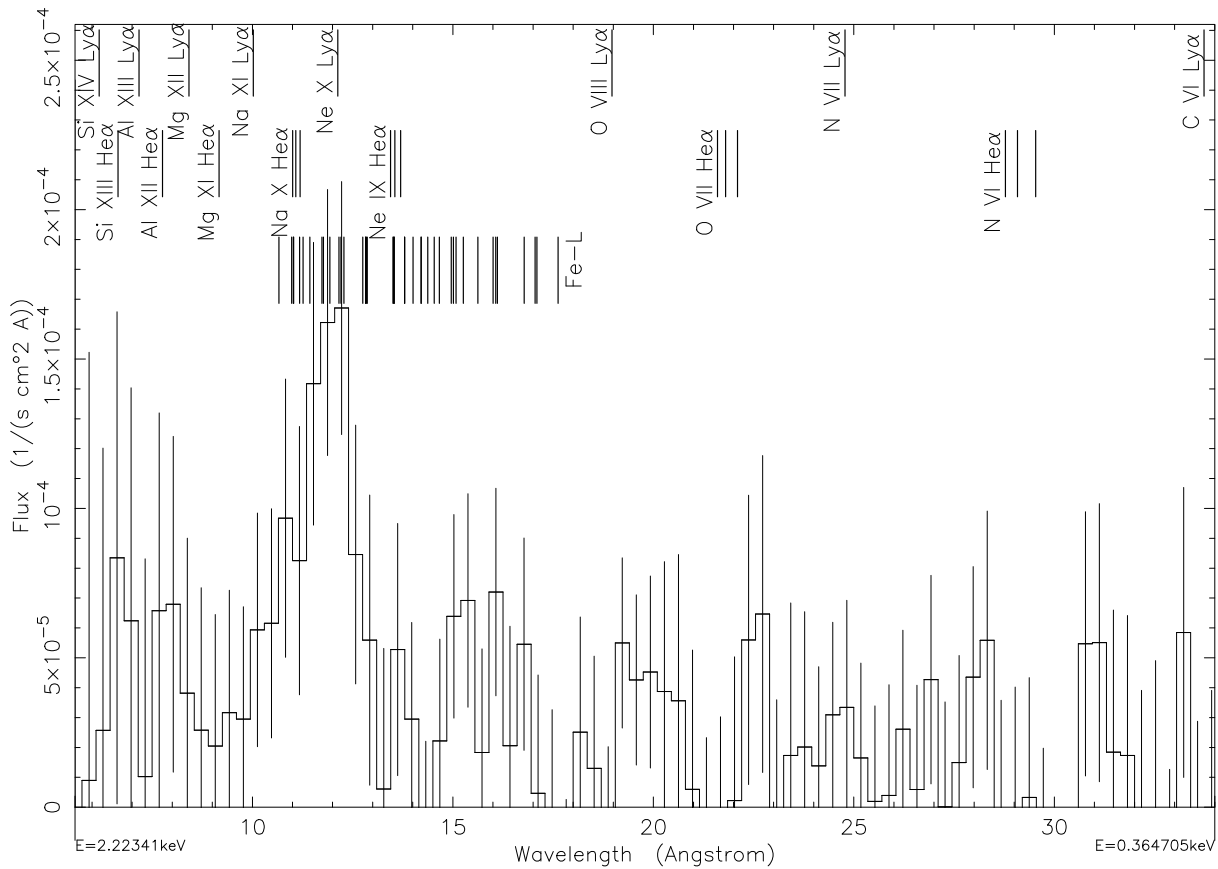}}
\caption{\label{fig:RGS}
The fluxed RGS spectrum of MKW4 (background-subtracted, RGS 1 \& 2,
orders 1 \& 2) in counts cm$^{-2}$ s$^{-1}$ \AA$^{-1}$. The
identifications of bright hydrogen-like and helium-like emission lines
are indicated, together with lines from the Fe-L series. Error bars are an
estimate of the statistical error on each bin, based on the square root of
the number of counts. 
}
\end{figure*}

First- and second-order spectra for RGS1 and RGS2 were extracted from
the early (S001/S002) and later (S004/S005) exposures (i.e. 8 spectra
in total). It is possible, before attempting any detailed spectral
fitting, to combine the spectra, together with the spectral response
matrices, into a single spectrum (this is useful for non-analysis
purposes to increase the signal-to-noise and to detect faint spectral
features). A single fluxed spectrum, in units of photons cm$^{-2}$
s$^{-1}$\AA$^{-1}$, was produced from the eight RGS spectra and is
shown in Figure~\ref{fig:RGS}.

Though we are not able to perform quantitative spectral fits to the
fluxed spectrum shown in Figure~\ref{fig:RGS}, some conclusions can be drawn
directly from the figure.

Ly$\alpha$ emission lines of hydrogenic charge states of a number of the
abundant low Z elements, particularly O and Ne (though some contribution
from Fe is likely here), are seen in the RGS spectrum, as are helium-like
charge states of a number of elements. Also present are neon-like and
fluorine-like charge states of iron (\Fexiixiii) and it appears that
emission from much of the Fe$-$L series is present (especially around 12\AA
[$\sim$1\,keV], where Ne may also contribute). The fact that lines such as
\Mgxii\ exist together with lines from much of the Fe$-$L series points
strongly towards emission from gas over a range of temperatures, rather
than emission from a single temperature gas. The presence of the \Ovii\ 
line suggests that some fraction of the gas has a temperature as low as
0.5~keV, but our data do not permit the determination of the amount of gas
involved. From the EPIC results, in particular the CEVMKL fit which shows a
sharp decline in emission at low temperatures, we assume that the amount
must be small.

It was expected, given the fact that the MKW4 emission covers nearly
all the RGS detector (hampering the extraction of reasonable
background spectra), and that the source is not especially bright,
that any detailed spectral modeling of the RGS data would be
difficult.

In order to perform detailed spectral analysis, separate response
files for each instrument and order (i.e. for each separate spectrum
created) were generated using \textsc{rgsrmfgen-0.50.2} (within rgsproc),
using, as is recommended, a large number (4000) of energy bins. These
responses were then attached to their relevant spectra and the spectra
were individually grouped into \xspec-usable spectra with a minimum of
20 counts per bin, a bin typically spanning $\sim0.04$\,\AA.

The eight spectra (two observations, RGS 1 \& 2, orders 1 \& 2) were
fitted simultaneously using a number of spectral models. In each model
a redshift for MKW4 of $z=0.02$ was used. It is very useful to combine
the data in this manner as gaps in the data due to non-functioning
CCDs (specifically RGS1 CCD 7 and RGS2 CCD 4) can be filled with data
from the other instrument and order(s).

Though a number of the ionization states of the Fe$-$L series appear
to exist, indicating that the gas may be non-isothermal, we were only
able to justify the use a one-component mekal model. Freezing the
hydrogen column at the Galactic value yielded a best fit temperature
of 1.04$-$1.41\,keV (1$\sigma$ confidence interval), and a best fit
metallicity of 0.27$-$0.62 solar. Letting the column free gave a best
fit value of (9.8$-$19.3)$\times10^{20}$\,cm$^{-2}$, with a
temperature of 1.02$-$1.33\,keV and metallicity of 0.19$-$0.49 solar.

For comparison with the EPIC fits we tested more complex models, but it
should be noted that these produced no significant improvement in the
quality of the fit, and often had poorly constrained parameters. Allowing
individual elemental abundances to vary did not significantly improve the
fits, and the abundances were essentially unconstrained. Fitting a
two-temperature MEKAL model with abundances tied resulted in fits very
similar to the single-temperature model, with the harder component making
little contribution.  A CEVMKL model fit gave a similar power-law slope to
that seen in the EPIC fits, but with unphysically high abundances. The
MKCFLOW model was more successful in fitting the RGS data than the EPIC,
but the parameters were effectively unconstrained, and error limits could
not be calculated. The minimum temperature found was $\sim$0.35 keV with
the hydrogen column frozen at the galactic value, and $\sim$0.1 keV with
\NH\ allowed to vary. The best fit metal abundance was approximately solar.
However, we stress that these fits are only included for comparison, as a
the statistics are such that we cannot justify the use of models more
complex than a single-temperature plasma model.

\section{Three-dimensional halo properties}

\begin{figure*}
\vspace{-1.5cm}
\centerline{\epsfig{file=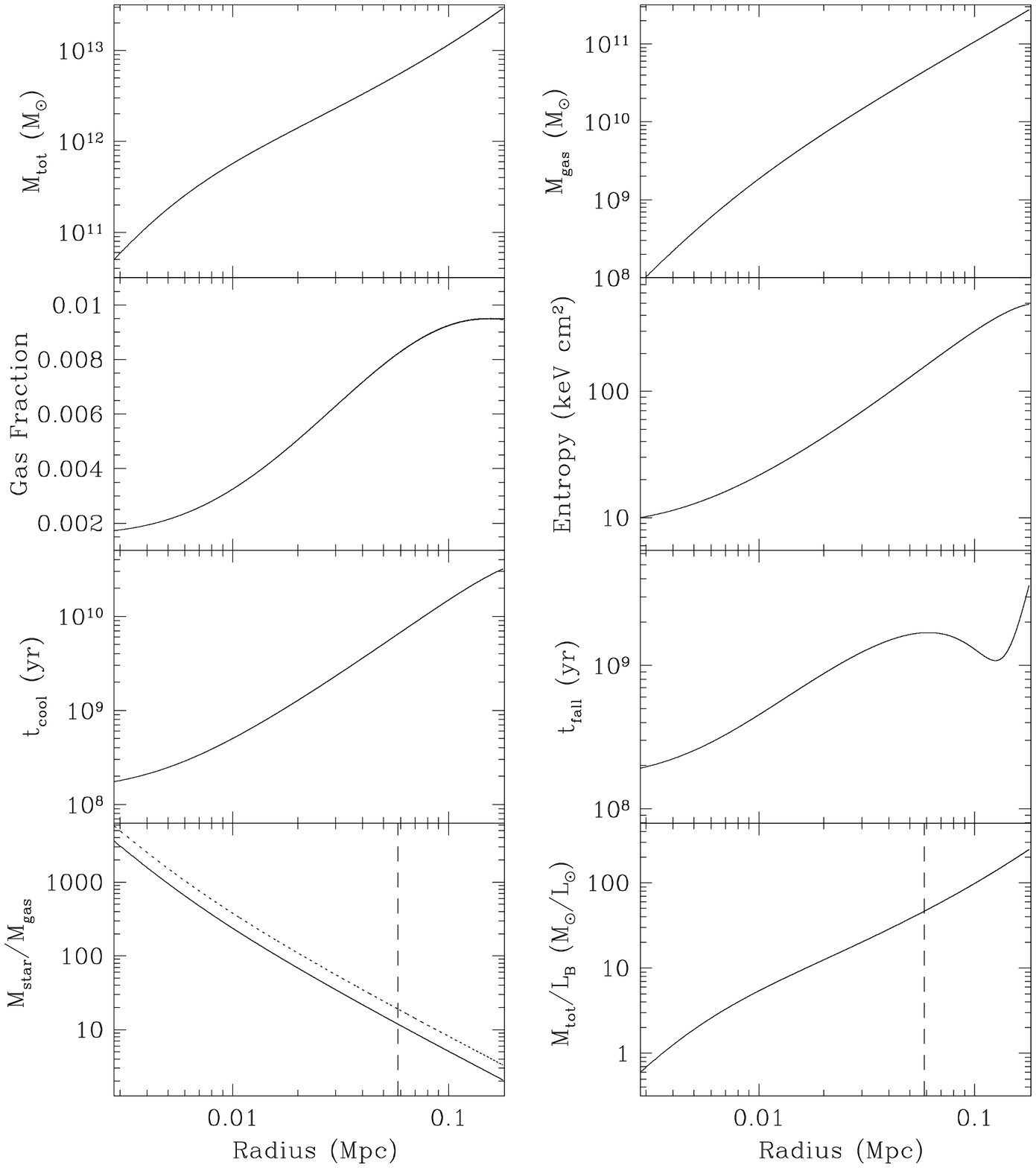,width=6in}}
\vspace{-2.5cm}
\caption{\label{fig:entropy} Radial profiles of total mass, gas mass, gas
  fraction, entropy, cooling time, time taken to fall to current
  temperature, mass-to-light ratio and stellar to gas mass ratio for the
  inner regions of the MKW 4 group. In the last two plots, dashed lines
  indicate the limit of the optical surface brightness profile. The solid
  (dotted) line in the final plot shows the profile assuming a stellar
  mass-to-light ratio of 5 (8) \ML.}
\end{figure*}

Using software provided by S. Helsdon, we used the surface brightness fit
derived in Section~\ref{sec:image} and the projected
radial temperature profile shown in Figure~\ref{fig:Tprof} to
calculate quantities such as gas mass, total mass, gas cooling time and gas
entropy for the inner region of the group. In principle we would wish to
use a deprojected temperature profile, but in this case the deprojected and
projected temperature profiles are similar in the core and the projected
profile has more stable fits at large radius. The software we use infers the
gas density profile based on the fitted surface brightness profile and a
model description of the temperature profile. In this case we model the
temperature profile using a quadratic, which is not physically meaningful
but gives an accurate estimate of the shape of the distribution. The
density profile is normalised by comparison with the X--ray luminosity of the
region in question, and scaled appropriately. Given this density profile we
can use the well known equation for hydrostatic equilibrium, to calculate the total mass at any given radius. 

\begin{equation}
M_{tot}(<r) = -\frac{kTr}{\mu m_p G}\left(\frac{d{\rm ln}\rho_{gas}}{d{\rm
      ln}r}+\frac{d{\rm ln}T}{d{\rm ln}r}\right),
\end{equation}

Entropy is simply defined
as S = T/n$_e^{\frac{2}{3}}$, where T is the temperature in keV and n$_e$
is the electron density of the gas. Figure~\ref{fig:entropy} shows
radial profiles for these parameters, as well as gas fraction. 


From the cooling time and temperature, we can estimate the rate of cooling
at a given radius. This can be used to estimate the timescale on which an
isothermal halo, allowed to cool, would assume the temperature profile
which we observe. The profile of $t_{fall}$ shows the result of this
calculation, where 

\begin{equation}
t_{fall} = \frac{T_{max}-T(r)}{\frac{dT}{dt}},
\end{equation}

$T$ is temperature and $t$ is time. In this case we chose to set
$T_{max}$=3 keV, the approximate mean temperature in the outer bins of our
temperature profile. We note that at all radii $t_{fall}$ is considerably
less than 7.5~Gyr, the approximate time since the last major merger of the
dominant galaxy.

Using the optical surface brightness profile of NGC~4073 measured by
\citet{MorbeyMorris83}, we have estimated the mass-to-light profile of the
inner part of MKW~4, and the stellar mass contribution. The mass-to-light
ratio is $\sim$50 \ML\ at the limit of the optical measurements. To
determine the mass of stars in NGC~4073 we must assume a stellar
mass-to-light ratio. We use a value of 5 \ML, but also show a profile for a
value of 8 \ML. In either case, the mass in stars clearly dominates over
the mass in gas to large radii. Within the central $\sim$10 kpc, we note
that the mass-to-light profile of the group falls below the ratio we have
assumed for stars. At this scale the temperature and surface brightness
structure of the group is resolved, so the problem is real, and suggests
that we are underestimating the mass in the core. One possible reason for
this is that the gas in the core may not be in hydrostatic equilibrium,
rendering our assumptions invalid. The relatively poor spectral fits
achieved in the core may also provide an explanation, if our measured
temperatures do not accurately represent the multi-phase gas in this
region.

\section{Discussion}
\subsection{Imaging analysis}
The halo of MKW 4 is not well described by an elliptical $\beta$-model
within the region we have studied. The emission shows signs of structure
along a NW-SE axis, with a bulge in the emission profile to the SE and
excess extended emission to the NW. This does not seem to be an ``edge'',
such as the structures associated with cold fronts in more massive clusters
\citep[\textit{e.g.,}][]{Markevitchetal00}. In this case the decrease in
emission is not sharp, with the bulge occuring between 35\arcs\ and
80\arcs, suggesting that we are not looking at an unresolved sharp-edged
feature. It does not correspond to any rapid change in hardness or fitted
temperature, at least at the resolution we can achieve with these data. If
there were large scale motions in the gas, we might expect to see sharp
features at the boundary of the region in motion, and the lack of such
features suggests that any motion is relatively minor, and that the halo is
likely to be close to hydrostatic equilibrium. One possible cause of large
scale gas movement would be the merger of a small subgroup, but this would
have to have taken place some time ago for the effects to have died down to
the current level.  The kinematic data and morphology segregation of the
galaxy population also argue against any recent merger, suggesting instead
that this system has not been disturbed for a significant period of time.

The surface brightness profile of the NE and SW quadrants is similar and
can be modeled with a single $\beta$-model.  We find that a model with
\Bfit$\sim$0.45 is a good fit to the data. This value of \Bfit\ is fairly
typical for galaxy groups of the mass and temperature range in which MKW 4
lies \citep{Sandersonetal03}, and compares well with a fit of the inner
6\arcm\ of the group using \rosat\ data \citep{OPC03}.
\citet{Helsdonponman00}, again using the \rosat\ data but with a much
larger region of interest (radius 18\arcm), find that the best fit to the
surface brightness distribution is a two component model. The more compact
component has \Bfit=0.73$\pm$0.06, while the extended component has
\Bfit=0.46$\pm$0.01. There is an offset between the two components, leading
to the classification of the halo as bimodal. Our region of interest is
much smaller than theirs (7.75\arcm\ compared to 18\arcm), so we would not
necessarily expect to accurately model the extended component of their fit.
The difference between their central component and our fit may well be
caused to some extent by the halo structure which we have observed and
excluded. If the compact component of their fit is required to partially
model the halo structure, this would suggest that we should compare our
model only to the more extended component of theirs. In this case the two
models agree reasonably well. A single beta model fit to the \rosat\ data
gave \Bfit=0.43, even closer to our best fit. We conclude that our fit is a
reasonable model of the structure-free portion of the inner halo, but note
that the \rosat\ fit is likely to be more accurate at large radii, owing to
the larger field of view.

\subsection{Spectral analysis}
The spectral fits to the PN and MOS data for the group halo show strong
signs that the gas is not well described by a single temperature model
within $\sim$40 kpc. The existence of low and high energy line emission in
the RGS spectra supports this. We do not obtain a statistically acceptable
fit to the EPIC spectra with any of the models we have used. Some degree of
error is to be expected, arising from inaccuracies or incompleteness in the
spectral modeling codes and instrumental responses. There is also the
question of background subtraction. The technique we have used to generate
our background spectra should provide us with the most accurate estimate of
the background available. However, as the process involves scaling and
combining a number of spectra from different sources, it is possible that
errors could occur and that these could affect our fits. The relatively low
level of the background suggests that this is most likely to be a problem
in the outer regions of the halo. In the central regions, we tested the
effect of varying the background normalisation is various of our fits, and
found that it generally has little effect. For example, best fit parameters
were not found to change by more than 1\% for fits to the five inner bins
of our projected radial profile when the then background was varied by
$\pm$20\%. It therefore seems unlikely that errors in the background are a
significant source of inaccuracy in the spectra fits. We conclude that our
models provide a reasonable approximation of the state of the gas, but the
multi-phase structure of the inner halo and (to some extent) the errors
associated with the calibration of instruments and spectral models prevents
us from truly modelling the the gas accurately.

The RGS instruments are sensitive mainly to the emission from the central
bright core of MKW~4. Although there will always be some emission from
large radii in the spectrum, we estimate that $\sim$90\% of the counts in
our extraction region are from the central 2\arcm\ of MKW~4. The RGS
spectral fitting results in terms of gas temperature and metallicity are
approximately consistent (to within $\sim$1$\sigma$) with the EPIC results
from the innermost regions (as would have been expected), the fact that the
RGS data yield a slightly lower temperature and metallicity than the EPIC
data is very suggestive of problems with the extraction of reasonably
representative background RGS spectra. Even at the edge of the RGS chips,
it appears that there is sufficient very low-surface brightness emission
from MKW4 to contaminate the extracted background spectra significantly.
The background spectra have a component of hot (and metal-rich) cluster
emission within them, such that the temperatures and metallicities obtained
from the spectral fitting of the bright source, using these
cluster-contaminated background spectra, are lower than for EPIC. Indeed,
experimenting with different values of the percentage of the PSF excluded
to create the background spectra showed that, as the background extraction
region moved closer to the edges of the detector (and into regions of less
cluster-contaminated background), the spectral fitting results tended to
and matched those obtained with EPIC for the innermost annuli. As a
downside, of course, such a procedure led to background spectra with
progressively smaller signal-to-noise ratios, such that the errors on the
spectral fitting results became unreasonably large. The results presented
here lie somewhere in between, and use a background extraction where the
background spectra statistics are still reasonable.

The most successful fits to the EPIC spectra of the central 2\arcm\ region
region are the 'two-temperature' APEC+Powerlaw and multi-temperature CEVMKL
models. Again, neither of these models is formally statistically
acceptable, but this may be caused by calibration and model inaccuracies
rather than indicating that the models are not good descriptions of the
emission. The difference in abundance between the two demonstrates the
difficulty of accurately modelling the system using a single
spectrum. Clearly the radial spectral profiles are more reliable in this
regard. Of the two, the APEC+Powerlaw appears to be the less likely from
a physical perspective. The index of the powerlaw is not particularly well
constrained, and seems more likely to be representing some high temperature
gas component than to be a product of an AGN or population of discrete
sources. The surface brightness models do not require a central point
source and there is no detection of AGN activity at other wavelengths. As a
further test we have looked for variability in the emission within the
central 25\arcs\ of NGC 4073, using light curves for each camera with a bin
size of 100s. The PN and MOS 1 cameras show no evidence of variations in
count rate larger than 3$\sigma$ from the mean. The MOS 2 camera has 3 bins
with count rates greater than the 3$\sigma$ limit, but in general emission
from the central region is fairly constant. If we force a MEKAL+powerlaw
fit to the two central bins of the temperature profile (see
Fig.~\ref{fig:Tprof}) we find a flux of 2.58$\times$10$^{-13}$
erg~s$^{-1}$~cm$^{-2}$ (0.2-8.0 keV) for the powerlaw component. This can
be considered as an upper limit on the luminosity of any AGN of log \Lx\ 
$<$ 41.29 \ergps. For reference, the flux [log \Lx] contribution from the
thermal component in the same region is 2.90$\times$10$^{-12}$
erg~s$^{-1}$~cm$^{-2}$ [42.35 erg~s$^{-1}$] (0.2-8.0 keV). The contribution
from X-ray binaries can be estimated by comparison with Chandra results
from other early-type galaxies. The \LxbLb\ ratios for NGC~720, NGC~4697
and NGC~1553 are 8.1, 7.5 and 7.2 erg~s$^{-1}$~\LBsol$^{-1}$ respectively
\citep{Jeltemaetal03}, leading us to derive an X-ray binary contribution of
Log \Lxb\ = 40.93-40.98 \ergps\ for NGC~4073. A more conservative upper
limit on the luminosity of any central source would therefore be Log \Lx\ 
$<$ 41.0 \ergps. As a further check on the existence of a central AGN, we
tried fitting a surface brightness model to the data including a central
point source with this luminosity. Such a fit is poor, with strong
residuals in the core. The fit can be improved if the point source
normalisation is allowed to fall by $\sim$30\%, but the residual map
suggests that the fit is still worse than that produced by a single beta model.

The CEVMKL model seems a more reasonable choice, suggesting that the halo
contains gas at a range of temperatures. However, it is a poorer fit to the
data, and the metal abundances required are high compared to the other
models and to previous studies with other instruments. The high value of
$\alpha$ found for this model suggests that there is very little emission
at low temperatures. Some confirmation of this can be drawn from the
MKCFLOW and VMCFLOW models, which although they are poor fits to the data,
do not find emission from gas cooling out of the X--ray regime. The poor
results achieved with these models strongly suggest that although the halo
contains gas at a range of temperatures, it is not well modeled as a steady
state constant pressure cooling flow.

\subsection{Abundances profiles and metal masses}
The profiles shown in Figure~\ref{fig:Tprof} suggest a strong variation in
either abundance or hydrogen column near the core of the group. NGC 4073
has a \Dtf\ radius of 1.41\arcm, or $\sim$32.5 kpc at our assumed distance.
This means that the central three bins include emission from NGC 4073. The
increase in hydrogen column is a factor of 2 or 3 times the measured
galactic value, except in the central bin, where it rises to 4 or 5 times
galactic. Such a large increase in \NH\ arouses some suspicion, given that
we necessarily use overly simplistic models to fit gas which is likely
multiphase, with gradients in both temperature and abundance. The excess
above the galactic column in the four central bins of our projected profile
translates to $\sim$1.3$\times$10$^{10}$ \Msol\ of hydrogen in the central
40 kpc of the system. NGC~4073 has no detected 21~cm emission
\citep{HuchtmeierRichter89}, and the upper limit on the mass of neutral
hydrogen in the system is 3.55$\times$10$^9$ \Msol \citep{Robertsetal91}.
This is relatively similar to our result, but even taking the 90\% lower
limits on our measured \NH\, we find a mass three times higher than the
radio-based upper limit. This strongly suggests our \NH\ measurements are
overestimating the true intrinsic absorption. The fact that the galactic
value provides a good fit at higher radii supports this conclusion. Holding
the hydrogen column fixed at the galactic level, we find a strong increase
in metal abundance in the core. This increase is also quite dramatic, with
average abundance rising by a factor of 2.  Previous work using \asca\ has
shown this increase in abundance in the core \citep{Finoguenovetal00}.
Comparison between our radial fits to individual elements, and those
derived from the \asca\ data shows good agreement at all radii, except
possibly in the central bin, where we find Si and S abundances somewhat in
excess of those found using \asca. If we use finer binning than that shown
in Figure~\ref{fig:Metals}, we find that the Iron abundance within 15\arcs\ 
also significantly exceeds the central value derived from the \asca\ data.
The poor spatial resolution of \asca\ necessitated larger radial bins than
we have used, so that the two central bins used in Figure~\ref{fig:Metals}
are equivalent to the central bin of the \asca\ study. It seems clear that
the high central abundances we see in the \xmm\ data were effectively
smoothed out by the poorer \asca\ spatial resolution.

In each radial bin we can use the gas mass and the abundances of Fe, Si and
S to calculate metal masses for each element. For the two innermost bins in
Figure~\ref{fig:Metals} we find the masses given in Table~\ref{tab:Metals}.
These values can be used to calculate the relative contributions to
enrichment of the ISM by SNIa and SNII. We assume yields of Si and Fe for
type II supernovae of y$_{Fe}$=0.07 \Msol\ and y$_{Si}$=0.133 \Msol\ 
\citep{Finoguenovetal00}, and from type Ia supernovae y$_{Fe}$=0.744 \Msol\ 
and y$_{Si}$=0.158 \Msol\ \citep{Thielemannetal93}. We estimate that
$\sim$49\% and $\sim$50\% of the iron within these regions is injected by
SNIa, in fairly good agreement with the peak values calculated based on the
\asca\ data. 

\begin{table}
\centerline{
\begin{tabular}{l|ccc}
Bin & \multicolumn{3}{c}{Mass (\Msol)} \\
 & Fe & Si & S \\
\hline
1 & 6.4$\times$10$^5$ & 7.2$\times$10$^5$ & 3.42$\times$10$^5$ \\
2 & 3.43$\times$10$^6$ & 3.64$\times$10$^6$ & 1.02$\times$10$^6$ \\
\end{tabular}
}
\caption{\label{tab:Metals} Mass (in solar units) of Fe, Si and S in the
  two innermost bins shown in Figure~\ref{fig:Metals}}
\end{table}

\subsection{Comparison with other systems}
Several other groups and clusters of comparable temperature have been
observed with the \xmm\ RGS, in general with considerably higher
statistics. The best examples are NGC~4636 \citep{Xuetal02}, NGC~5044
\citep{Tamuraetal03} and M~87 in Virgo \citep{Sakelliouetal02}. These
objects all show evidence of multi-phase gas in their central regions, but
in each case there is a lower limit to the observed temperature range. Gas
below this limit is either nonexistent, or present only in very small
quantities. The minimum (and maximum) temperatures observed for these
objects are 0.6 (1.1), 0.6 (1.8) and 0.53 (0.71) keV for NGC~5044, M~87 and
NGC~4636 respectively. The minimum temperatures in each case are quite
similar to each other, and to the minimum temperature of $\sim$0.5 keV
which we infer from the presence of the \Ovii\ line in our RGS spectra. The
variation in maximum temperatures is larger, but NGC~5044 has a similar
maximum temperature to that which we derive from our RGS data ($\sim$1.2
keV). While the maximum temperature seems likely to be related to the
temperature of the large scale halos surrounding each galaxy, and therefore
to the mass of each group or cluster, the similarity in minimum
temperatures is intriguing. Clearly any models intending to explain the
apparent absence of traditional cooling flows in these objects must be able
to reproduce this limiting temperature.

\xmm\ observations of similar systems also show abundance gradients, which
are in some ways similar to those in MKW~4. A detailed analysis of the EPIC
data for M87 reveals a strong decline in the abundances of Fe, Si, S, Ca
and Ar with radius \citep{Matsushitaetal03}, with abundances falling from
$\sim$1.5 \Zsol\ in the core to \lesssim0.5 \Zsol\ at a radius of 10\arcm.
Only oxygen is relatively constant at all radii. The high central
abundances of Si and S are quite comparable to those we observe in MKW~4,
but it is notable that in M87 Si, S and Fe share similar radial abundance
profiles and have similar abundances at all radii. 

An observation of NGC~5044 with the RGS \citep{Tamuraetal03} shows some
evidence of Si abundance in excess of Fe, though it is consistent with a
solar ratio when the estimated systematic errors are accounted for.
\citet{Buoteetal03a}, using spectra from the EPIC instruments and the
\chandra\ ACIS-S3 find similar results. Single temperature fits show Si and
S to be marginally more abundant than Fe in the core, but 2-temperature
fits give significantly increased abundances of Fe and Si, with Fe the most
abundant of the three \citep{Buoteetal03b}. The 2-temperature models
provide a significantly better fit in the central $\sim$30~kpc of the
group, and the resulting abundances lead to the conclusion that 70-80\% of
the Fe mass in the central 100~kpc of NGC~5044 has been produced by
SNIa. However, this analysis assumes the abundance ratios of
\citet{GrevesseSauval98}, which give rather higher Fe abundances than the
ratios we use \citep{AndersGrevesse79}. If we assume that the effect of
this difference is a factor of $\sim$1.4 in Fe abundance, then our results
are fairly consistent with those found for both M87 and NGC~5044.

\subsection{Mass, cooling time and entropy}
We use an estimate of \Rth\ (the radius at which the system has a density
of 200 times the critical density of the universe, approximately equal to
the virial radius) based on fits to \rosat\ data \citep{Helsdonponman00}.
This estimate, \Rth=973 \hsf$^{-1}$ kpc using our chosen \Ho, assumes the
temperature profile to be isothermal, which we know to be inaccurate. This
value compares well with the value we would derive from the work of
\citet{NavarroFW95} (\Rth=935 \hsf$^{-1}$ kpc). However,
\citet{Sandersonetal03} point out that \citet{NavarroFW95} do not include
preheating in their models, and that estimates based on \rosat\ and \asca\ 
data suggest that their models overestimate \Rth\ for low mass systems such
as MKW~4. \citet{Sandersonetal03} estimate \Rth=786 \hsf$^{-1}$ kpc for
this system. Unfortunately we cannot estimate \Rth\ from our fits since
they only model the data well in the inner part of the group. We therefore
assume the \Rth\ value of \citet{Helsdonponman00}, which means that our
profiles extend to $\sim$19\% of the virial radius, but note that the
\citet{Sandersonetal03} estimate would mean that they extend slightly
further, to $\sim$23\%.

The total mass within this radius seems quite comparable to that found for
systems of similar X--ray temperature \citep{Nevalainenetal00}, but the gas
mass appears at first glance somewhat low, with the gas fraction of the
group only reaching $\sim$1\%. However, comparison with other cool systems
\citep{Sandersonetal03} suggests that this value may not be unrealistic,
given the low fraction of the virial radius to which we can measure. The
flattening of the gas fraction profile at large radii is caused by the
flattening and turnover of the temperature profile at these radii. Although
we cannot measure the temperature profile to larger radii, the \rosat\ 
analysis of \citet{Helsdonponman00} extends it to $\sim$420 kpc. The
turnover of the temperature profile is observed in the \rosat\ data, but at
larger radii the profile flattens again, apparently becoming roughly
isothermal with a temperature similar to that we observe in the core. Our
model of the temperature profile clearly cannot be extrapolated, but the
\rosat\ profile suggests that gas fraction should begin rising again just
outside our region of interest. We find that the gas entropy at
0.1$\times$\Rth\ is in good agreement with previous studies of groups and
clusters \citep{Lloyd-Daviesetal00}, when corrected for differences in
assumed \Ho.

As has been observed in other galaxy groups
\citep{Ponmanetal03,Mushotzkyetal03}, there is no central isentropic core.
Some models of entropy injection in galaxy groups and clusters predict a
flattening of the entropy profile in the cores of the systems
\citep[\textit{e.g.,}][]{Muaonwongetal01,TozziNorman01}, and while this is
observed in some higher mass systems \citep[\textit{e.g.,}][]{Davidetal01},
it is clearly not the case here. The cooling time is fairly short in the
core of NGC 4073, but rises to $\sim$10$^9$ yr at the \Dtf\ radius (the
radius at which the average B-band surface brightness has dropped to 25
mag/arcsecond$^{2}$), and is longer than a Hubble time at the outer edge of
our measurements. The short cooling time in the core, and short timescale
required for the halo to have cooled to its current state suggests we are
not observing a long-established cooling flow. It seems more likely that
the gas is cooling, but that a large scale flow has not yet had sufficient
time to develop. However, the relatively short timescale on which the gas
is able to cool to its current state, $t_{fall}$, suggests that some other
source of energy is required to prevent the gas cooling further, a source
which was active in the more recent past. The value of $t_{fall}$ in the
core indicates that this energy source must have been active
$\sim$2$\times$10$^8$ yr ago. This timescale is similar to that estimated
for AGN driven by cooling flows \citep{BinneyTabor95}. We note that at the
minimum radius for which we calculate \tcool, 2 kpc (equivalent to the
mid-point of our innermost surface brightness bin), it is
1.17$\times$10$^8$ yr, very similar to the minimum value of $t_{fall}$. The
profiles in Figure~\ref{fig:entropy} have a minimum radius equal to the
midpoint of our innermost temperature bin, within which we cannot be
certain our profiles match the true temperature structure.

\subsection{Halo evolution models}
The most obvious reason for the observed level of enrichment by SNIa in the
core of the group is that the gas has been there for some time, allowing
the metal abundance to build up. One possibility is that instead of a group
with a steady decrease in temperature towards the core, we are viewing an
essentially isothermal group halo with a cooler galaxy halo embedded within
it. The cooler region is certainly centred on NGC 4073, and such a sizable
galaxy might be expected to be capable of producing a large gaseous halo
through stellar mass loss and accretion. The temperature in the core,
$\sim$1.4 keV, would be quite high for an elliptical galaxy, but not
impossibly so, particularly when contamination by group emission is taken
into account. The high metallicity of the core, and suggestion of increased
SNIa enrichment within the galaxy is a strong argument for a galaxy halo on
some scale. The strongest argument against this explanation is that
spectral fits using two temperature models do not seem as good as we might
expect if this were the case. The relatively poor constraints on the hotter
component in the APEC+APEC and APEC+powerlaw fits, the moderate success of
the CEVMKL fit and the existence of low and high energy lines as well as
lines from the Fe-L series in the RGS data all argue for multi-temperature
gas, rather than two separate components.

An alternative is that there is a cooling flow operating in the core of the
group, in which there are small scale inhomogeneities in the distribution
of metals \citep{MorrisFabian02}. This scenario could lead to apparent high
metallicities in the core while reducing the emission from gas at low
temperatures, as the high metallicity components would cool very rapidly at
low temperatures via line emission.  This model would predict a silicon
abundance in the inner core 1.5-2 times that of iron, and we observe a
ratio of $\sim$1.75. The model also predicts a central decrease in
abundance, as metal rich gas clumps in the core cool completely out of the
X-ray regime. Such abundance dips are seen in M87 \citep{Matsushitaetal03},
the Centaurus cluster \citep{SandersFabian02} and the NGC~5044 group
\citep{Buoteetal03b}, with scales of up to 15 kpc. The limits of resolution
for our abundance profiles are $\sim$6 kpc, and we do not see a definite
central dip. The two inner most bins of the deprojected abundance profile
are quite similar however, and the degree of error on each makes it
difficult to rule out such a dip. If there is a cooling flow in MKW 4 it
seems likely that it is small, affecting a region only a few tens of kpc
across, and this model may be a good description of it.

A third option is that the MKW~4 is a truly multi-temperature system, which
is prevented from developing into a cooling flow by some heating mechanism.
The lack of AGN activity in NGC 4073 (or clear indications of past
activity) is interesting, in that AGN outbursts have been suggested as a
method of reheating cooling gas \citep[\textit{e.g.}][]{Bohringeretal02}.
Modeling of AGN outbursts as a means of controlling cooling in Hydra A
\citep{KaiserBinney03} suggests that large scale cooling can be prevented
if the AGN produces bubbles in the gas involving $\sim$1\% of the halo. The
models also predict a duty cycle for a central AGN powered by infall of
cooling gas, of $\sim$200 Myr, very similar to the cooling time in the
central regions of the cluster.  MKW 4 is somewhat less dense and
considerably cooler than Hydra A, leading us to expect a shorter cooling
time and AGN duty cycle. We would also expect a similar fraction
(of order 1\%) of the gas halo to be involved in the bubbles produced by
the AGN heating. The lack of a detected AGN does not rule out this model
for MKW 4, nor does the fact that we do not see evidence of cavities in the
X--ray halo caused by old radio lobes.  If the AGN has been dormant for
some time, our field of view may be too small to detect any cavities, or
our observation may be too shallow.  However, as the galaxy population is
relaxed and morphologically segregated, with no significant substructure,
arguing strongly against any large merger event in the recent history of
the group, AGN activity would seem to be a promising candidate as a means
of preventing cooling in the core.

\section{Summary and Conclusions}
We have carried out an analysis of an \xmm\ observation of MKW~4,
studying the surface brightness, temperature and abundance structure of the
system, and using these results to infer the properties of the cluster
potential well and X--ray halo. The main results of the analysis can be
summarized as follows;

\begin{enumerate}
\item The surface brightness structure of the group suggests that it is not
  completely relaxed, but is nearly so. There is some structure in the halo
  along a NW-SE axis, which may indicate movement of gas in the halo, but
  no sharp edges or fronts. We conclude that any movement must therefore by
  weak, and that the group is near hydrostatic equilibrium.
\item The RGS spectra show that in the group core gas exists at a range of
  temperatures. The lowest temperature feature observed in these spectra is
  the \Ovii\ line. This suggests that there is a component of the gas in
  the core with a temperature of $\sim$0.5 keV, but no significant
  quantities of gas cooler than this.
\item Spectral imaging with the EPIC cameras shows that the group has a
  temperature profile which declines toward the core, and an abundance
  profile with a fairly strong central peak, in agreement with previous
  results from \rosat\ and \asca\
  \citep{Helsdonponman00,Finoguenovetal01}. In the central $\sim$40 kpc the
  halo is not well modeled by a single temperature plasma, suggesting
  either that we are not resolving the temperature structure in the core or
  that the gas is multi-phase.
\item From the abundances we measure we can estimate the numbers of SNII
  and SNIa required to enrich the intra-group medium to its current
  state. We estimate that $\sim$50\% of the Fe in the central 40 kpc was
  produced in SNIa. Our abundances compare well with those found in other
  objects of similar mass observed by \xmm\ and \chandra.
\item An analysis of the 3-dimensional properties of the system shows that
  it has a total mass similar to other systems of comparable temperature,
  and a gas entropy which agrees well with other systems. The gas fraction
  of the cluster is at rather low ($\sim$1\%), but still within the range
  expected for a system of this temperature. The cooling time in the group
  core is relatively short ($\sim$2$\times$10$^8$ yr), and is similar to
  the time we estimate the system would have taken to cool to its present
  state if the halo was once at a uniform temperature of 3~keV, equal to
  the peak temperature now observed.
\end{enumerate}

From these results we conclude that the group halo is a multi-temperature
system, but does not contain a large cooling flow, as indicated by the lack
of gas at temperatures below $\sim$0.5~keV. The high abundances observed in
the core suggest that the gas in the centre of the group has occupied its
current position for a reasonably long time, allowing enrichment by
supernovae. Although there is no direct evidence of AGN activity in the
dominant galaxy of the group, or of past activity, we consider it likely
that the group is periodically heated by a central AGN, probably fueled by
gas cooling out of the halo. At present the AGN is dormant, as the halo has
not cooled sufficiently to provide fuel for an outburst. Alternative
sources of heating, such as star formation or the merger of a smaller group
are ruled out by the optical data, which suggest that both group and
dominant galaxy are old and undisturbed. We therefore conclude that MKW~4
is most likely to be a system governed by feedback between cooling in the
group halo and heating by an AGN in its dominant galaxy.

\vspace{1cm}
\noindent{\textbf{Acknowledgments}\\
  The Authors would like to thank B. Maughan for the use of his software
  and advice on XMM analysis, and S. Helsdon for providing his 3-d
  properties script. We would also like to thank G. Mamon, J.~Kempner and
  A.~Sanderson for useful discussions and suggestions. Lastly we would like
  to thank the referee, A.~Lewis, for a thorough reading of the paper and
  numerous suggestions which have significantly improved it. This work made
  use of the Digitized Sky Survey, the NASA/IPAC Extragalactic Database,
  and Starlink facilities at the University of Birmingham. This research
  was supported in part by NASA grants NASA NAG5-10071 and NASA GO2-3186X.

\bibliographystyle{mn2e}
\bibliography{../paper}

\label{lastpage}

\end{document}